\documentclass{imsart}
\usepackage{mathrsfs}
\arxiv{math.PR/0000000}
\pubyear{\today}

\RequirePackage[authoryear]{natbib}
\usepackage{amsgen,amsmath,amstext,amsbsy,amsopn,amssymb,latexsym, amsthm}
\usepackage[usenames]{color}
\usepackage{array}
\usepackage{multirow}
\usepackage{graphicx}
\usepackage{comment}
\usepackage{xcolor}
\usepackage{booktabs}
\usepackage{float}
 \RequirePackage{subcaption}

\startlocaldefs

\newcommand{\var}[1]{\operatorname{Var}\left(#1\right)}
\newcommand{\cov}[1]{\operatorname{Cov}\left(#1\right)}

\newtheorem{lemma}{Lemma}

\newtheorem{theorem}{Theorem}

\def\bbK{\bbK}

\def\bX{\mathbf{X}}

\def\bx{\mathbf{x}}
\def\by{\mathbf{y}}

\def\1{{\mathbf 1}}

\def\bbK{\mathbb{K}}

\def\E{{\rm E}\,}

 \newcommand{\ignore}[1]{}
 
\newcommand{\wanjie}[1]{\textcolor{black}{#1}}

\def\mx{\mathbf{x}}
\def\my{\mathbf{y}}
\def\tmx{\tilde{\mathbf{X}}}
\def\tmy{\tilde{\mathbf{Y}}}
\def\E{{\mathrm{E}}}

\def\Eo{{\mathrm{E}}_1}
\def\are{\textsc{are}}
\def\var{{\mathrm{Var}}}
\def\cov{{\mathrm{Cov}}}

\newtheorem{lemx}{Lemma}

\endlocaldefs

\begin{document}

\begin{frontmatter}

\title
{Truncated Rank-Based Tests for Two-Part Models with Excessive Zeros and Applications to Microbiome Data}

\begin{aug}
		\author[A]{\fnms{Wanjie} \snm{Wang}\ead[label=e1]{wanjie.wang@nus.edu.sg}},  
		\author[B]{\fnms{Eric} \snm{Chen}\ead[label=e2]{}}
		\and
		\author[B]{\fnms{Hongzhe} \snm{Li}\ead[label=e2,mark]{hongzhe@upenn.edu}}

		
		\thankstext{t3}{This research was supported by NIH grants GM123056 and GM129781. Hongzhe Li is the corresponding author. 
		 }
		
		
		\address[A]{National University of Singapore,
		\printead{e1}}
		\address[B]{University of Pennsylvania,
			\printead{e2}}
	
	\end{aug}

		
		
		
		





\begin{abstract}
High-throughput sequencing technology allows us to test the compositional difference of bacteria in different populations. One important feature of human microbiome data is that it often includes a large number of zeros. Such data can be treated as being generated from a two-part model that includes a zero point-mass.  Motivated by  analysis of such non-negative data with excessive zeros, we introduce several truncated rank-based two-group and multi-group tests, including a  truncated rank-based Wilcoxon rank-sum test for two-group comparison and two truncated  Kruskal-Wallis tests for multi-group comparisons.    We  show both analytically through  asymptotic relative efficiency analysis and by simulations that the proposed tests have higher power than the standard rank-based tests in typical microbiome data settings, especially when the proportion of zeros in the data is high. The tests can also be applied to repeated  measurements of compositional data via  simple within-subject permutations. \wanjie{In a simple before-and-after treatment experiment, the within-subject permutation is similar to the paired rank test. However, the  proposed tests handle the excessive zeros, which leads to a better power. } We apply the tests to compare the microbiome compositions of healthy children and pediatric  Crohn's disease patients  and to assess the treatment effects on microbiome compositions. We identify several bacterial genera that are  missed by the standard rank-based tests.   
\end{abstract}

\begin{keyword}
Asymptotic relative efficiency; Differential abundance analysis; Two-part model; Truncation; 
\end{keyword}


\end{frontmatter}


\section{Introduction}\label{sec:intro}

The human microbiome  includes all microorganisms in various human body sites such as gut, skin and mouth. Gut microbiome has been shown to be associated with many human diseases, including  obesity, diabetes and inflammatory bowel disease  \citep{turnbaugh2006obesity,qin2012metagenome,manichanh2012gut}.   Two high-throughput sequencing based approaches, including 16S  ribosomal RNA (rRNA) sequencing and shotgun metagenomic sequencing, are commonly used in microbiome studies \citep{turnbaugh2007human, qin2010human}.
Bioinformatics methods are available for quantifying the microbial relative abundances based on such  sequencing data, which typically involve aligning the reads to some known database  or marker genes \citep{huson2007megan, segata2012metagenomic}. Since the DNA yielding materials are different across different samples, the resulting numbers of sequencing reads vary greatly from sample to sample. In order to make the microbial abundance comparable across samples, the abundance in read counts is usually normalized to the relative abundance of the bacteria observed, which results in high dimensional  compositional data.  Some of the most widely used metagenomic processing software such as MEGAN \citep{huson2007megan} and MetaPhlAn \citep{segata2012metagenomic} only outputs the relative abundances of the bacterial taxa. 

In microbiome studies, one is often interested in identifying the bacterial taxa such as genera or species that show different distributions between two or more conditions. One important feature of microbiome compositional data  is that the data include large clumps of zeros that represent the absence of the bacterial taxa in the samples, especially for those relatively rare taxa.  Zero observations can also result from under-sampling of the sequencing reads for rare taxa. As an example, Figure \ref{fig:heatmap}  shows the
heatmap of zeros   and the relative abundances for 26 healthy and normal samples  and samples from 85 Crohn's disease children for each of the 60 bacterial genera. These data were collected at the University of Pennsylvania \citep{IBD}. We are interested in identifying the bacterial genera that have different distributions between healthy and Crohn's disease patients.  We observed over 62.5\% of the observations are zero. 
 In addition, the compositional data are often  skewed, which makes parametric modeling of such data difficult and  tests based on parametric distributional assumptions   problematic. We present detailed analysis of this data set in Section \ref{sec:data}.
 
 Non-parametric tests, such as the Wilcoxon rank-sum test and Kruskal-Wallis test, can be applied to such data and are commonly used in analysis of microbiome data.  However, such rank-based tests tend to 
have low power because of the large number of ties from zero observations \citep{Lachenbruch1, Lachenbruch2, Hallstrom}.  A two-part test combining the square of a test statistic for comparison of the proportion of zeros and the square of an appropriate normal test such as   the Wilcoxon rank-sum to
compare the non-zero scores  was proposed and evaluated by \cite{Lachenbruch2}.  This two-part test  was recently applied to analysis of microbiome data \citep{Wagner}. 
\wanjie{However, the theory developed for Lachenbruch's 2 degree of freedoms  $\chi^2$ test  assumes that binomial test statistic and the test statistics for the continuous part to be independent, which only holds under the assumptions of independent errors of the binomial and continuous part of the distribution \citep{lachenbruch2002analysis}.   
However, such an assumption may not hold for the microbiome relative abundance data generated by sequencing since both can depend on the sequencing depth.  }

\begin{figure}[ht!]
	\centering
	\includegraphics[width=0.998\textwidth, height=0.35\textheight]{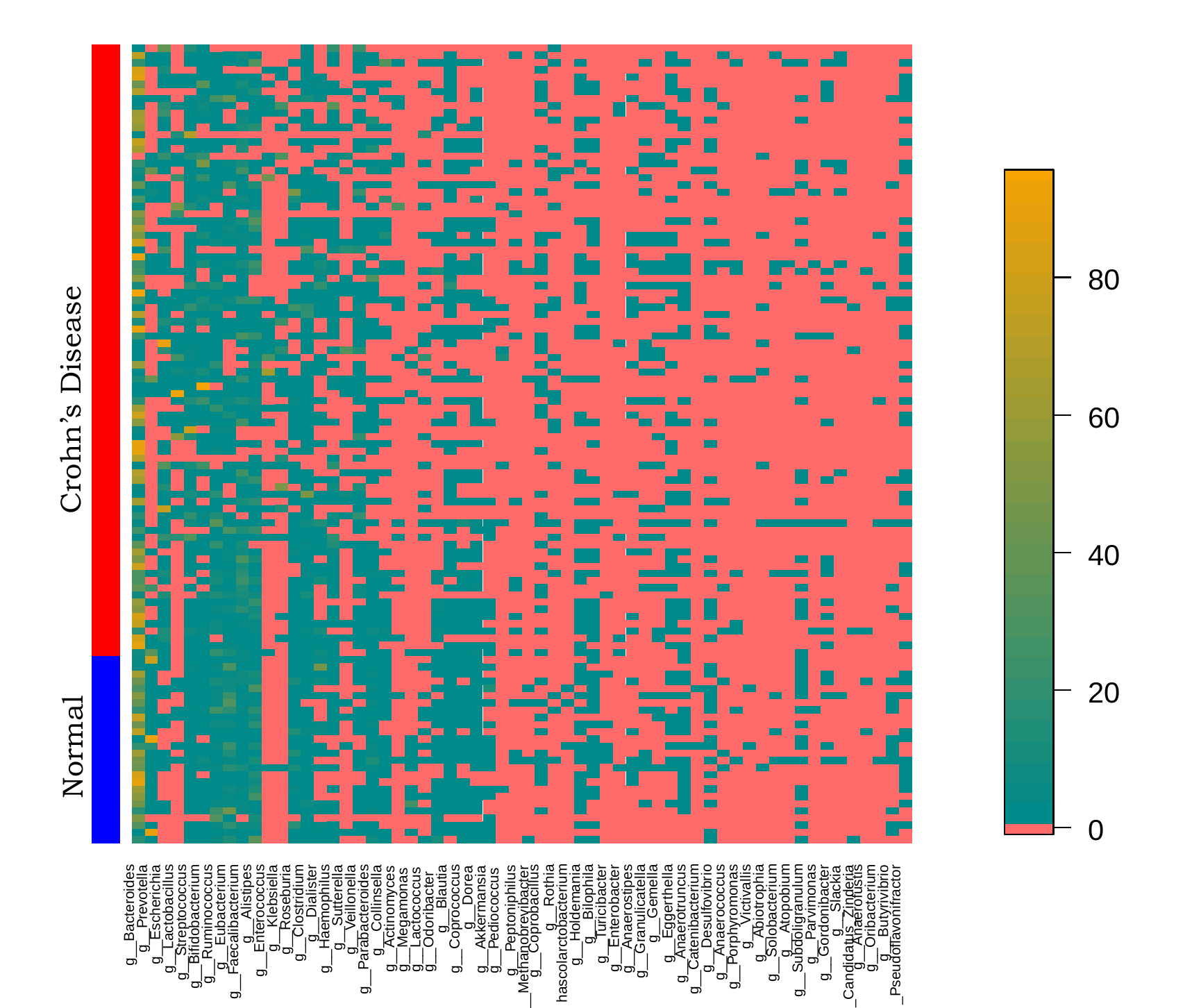}
	\caption{Heatmap of zeros   and the relative abundances of 60 bacterial genera for 26 healthy children and 85 patients with pediatric  Crohn's disease, showing over 62.5\% of the observations being zero \citep{IBD}. } 
	\label{fig:heatmap}
\end{figure}

To account for excessive  zeros in non-negative distributions,  \cite{Hallstrom} introduced a truncated Wilcoxon rank-sum test, where the Wilcoxon rank-sum test is performed after removing an
equal and maximal number of zeros  from each sample.   He showed that this test recovers much of the power loss from the standard application of the Wilcoxon test.  Compared with a directional 
modification of the two-part test proposed by Lachenbruch \citep{Lachenbruch1, Lachenbruch2}, the truncated Wilcoxon test has similar power when the non-zero relative abundances  are independent of the proportion of zeros. In addition,  the truncated Wilcoxon test  is relatively unaffected when the error terms of the two distributions  are dependent. \cite{Hallstrom} however only considered the two-sample test under the setting of equal sample sizes. 

In this paper, we assume that the data are generated from two-part models with point-mass at zero as one of the components. However, we do not make any distributional assumption on the continuous non-zero part. We  develop several rank-based tests  for general two-group comparison   with possible unequal sample sizes, and for  the multiple-group  comparisons with equal or unequal sample sizes.  Particularly, we extend the truncated  Wilcoxon rank-sum test of \cite{Hallstrom} to data with unequal sample sizes, and develop  a modified Kruskal-Wallis test to account for clumps of zeros for multiple-group comparisons  with equal sample sizes and unequal sample sizes, respectively. These new tests  are based on the idea of data truncation and asymptotic calculations and can effectively deal with the clumps of zeros in the data. The asymptotic null distributions of the tests are given. The key difficulty of deriving  such truncated rank-based tests is to calculate the variance of the test statistic under the null, which does not have a closed-form expression.  We instead use  asymptotic analysis to obtain approximations of the variance estimates.

In order to demonstrate the advantages of the proposed truncated rank-based tests,  we also derive the asymptotic relative efficiency of the proposed tests compared to commonly used Wilcoxon rank-sum test and  Kruskal-Wallis test when the data are generated from two-part models \citep{Lachenbruch2}.  We observe in our simulations  large gains in efficiency, especially when the proportions of zeros in the data are high.   These tests are rank-based, easy to calculate and provide new tools for identifying the  bacterial taxa with different distributions among different groups in human microbiome studies. We apply and compare the proposed tests by analyzing  a microbiome study conducted at the University of Pennsylvania \citep{IBD}, including  comparing  the gut microbiome difference between  healthy and pediatric  Crohn's disease patients and   assessing  the effects of treatment over time. 

\ignore{
The rest of the paper is organized as follows. In Section \ref{sec:wilcoxon}, we extend the truncated  Wilcoxon rank-sum test of \cite{Hallstrom} to data with unequal sample sizes. We then present in Section \ref{sec:kwequal}  and Section \ref{sec:kwdiff} a modified Kruskal-Wallis test to account for clumps of zeros for multiple-sample setting with equal sample sizes and unequal sample sizes, respectively.  For all these extensions, we derive the asymptotic relative efficiency compared to the standard rank-based tests. In Section \ref{sec:simu}, we present simulations to evaluate the proposed tests and to compare their  power. In Section \ref{sec:data}, we present results from an analysis of pediatric Crohn's disease microbiome study at the University of Pennsylvania to identify the bacterial genera that are associated with pediatric Crohn's disease and to identify the bacterial genera that change their abundances over time during the anti-TNF treatment. Finally, we give a brief summary and discussion in Section \ref{sec:diss}. Details of the mathematical derivations and proof of relevant lemmas are provided as  online Supplemental Materials. 
}

\section{A truncated Wilcoxon rank-sum test for data with  excessive zeros}\label{sec:wilcoxon}

\subsection{A truncated Wilcoxon rank-sum test}
Consider the two-sample setting where we have $N_1$ non-negative independent observations from population 1, $\mathbf{X}=(x(1),\cdots,x(N_1))$, and $N_2$ independent observations from population 2, $\mathbf{Y}=(y(1),\cdots,y(N_2))$,  where $x(i)$ (or $y(i)$) represents the relative abundance of a  bacterium in  the $i$th sample of group 1 (or 2). For  most of the bacteria, the data $\mathbf{X}$ and $\mathbf{Y}$ include many zeros, which represent absence of the bacterium in these samples or below detection limits. 
We are interested in testing whether these observations $\mathbf{X}$ and $\mathbf{Y}$ are from the same distribution, i.e., the hypothesis testing problem that 
\[
H_0: \,\, \mathbf{x} \sim F, \, \mathbf{y} \sim F \qquad vs \qquad H_1: \,\, \mathbf{x} \sim F, \, \mathbf{y} \sim G, F \neq G, 
\]
where $F$ and $G$ are both probability density functions with a proportion of zeros. We  consider the two-part model of \cite{Lachenbruch1}, which assumes that the data are generated from the following  distributions, 
\begin{equation}\label{AREmodel}
	x(i) \sim (1 - \theta_1) \delta_0 + \theta_1 f,\quad
	y(j) \sim (1 - \theta_2) \delta_0 + \theta_2 g, \quad 
	i = 1, \cdots, N_1, j = 1, \cdots, N_2,
\end{equation}
where \wanjie{$\theta_1$ (or $\theta_2$) is the probability of being non-zero in  population $1$ (or $2$), } 
$\delta_0$ is point mass at 0, and $f$ (or $g$) is the distribution for nonzero element in  the population 1 (or 2).  

Because of the excessive  zeros, the standard non-parametric Wilcoxon rank-sum test statistic is less effective. A truncated Wilcoxon rank-sum test statistic has been proposed by \cite{Hallstrom} for the case $N_1 = N_2 = N$. 
We first  extend his test to  the general setting where $N_1 \neq N_2$, and examine the asymptotic relative efficiency compared to the standard Wilcoxon test statistic.

Given $\mathbf{X}$ and $\mathbf{Y}$, denote $n_1$ (or $n_2$) as the number of  non-zero observations in $\mathbf{X}$ (or $\mathbf{Y}$) and 
let $p_1$ (or $p_2$) be the proportion of non-zero observations in $\mathbf{X}$ (or $\mathbf{Y}$), where  $p_1 = n_1/N_1$ (or $p_2 = n_2/N_2$). 
Let $p = \max\{p_1, p_2\}$. \wanjie{Let $\lfloor a \rfloor$ denote the largest integer that is smaller than $a$. } 
We rank the combined observations $\mathbf{X}\cup \mathbf{Y}$ from the largest to the smallest so that zeros have the highest  ranks, where  the tied measurements are given the average rank. 
For $\mathbf{X}$ (or $\mathbf{Y}$), we keep only $\lfloor{ p N_1 }\rfloor$ (or $\lfloor{ p N_2 }\rfloor$) observations with the smallest ranks, which  implies that the observations removed are all zeros. The truncated samples are denoted as $\tilde{\mathbf{X}}$ and $\tilde{\mathbf{Y}}$, respectively. 
Let $R$ denote the sum of the ranks of  all observations in $\mathbf{X}$. The Wilcoxon test statistic can be written as 
\begin{equation}\label{eqn:defW}
	T_{W} = \frac{S^2}{\mathrm{Var}[S]}, \quad \mbox{where } S = R - \frac{N_1 + N_2 + 1}{2} N_1,
\end{equation}
and $\mathrm{Var}[S]$ is the variance of $S$ under the null hypothesis. 

The same procedure can be applied to the truncated data $\tmx$ and $\tmy$. Rank the combined observations $\tmx \cup \tmy$, from the largest to the smallest, and let $r$ denote the sum of ranks of all observations in $\tmx$. 
We define the counterpart of $S$ as 
\begin{equation*}\label{smallS}
	s = r - \frac{\lfloor p(N_1 + N_2) \rfloor + 1}{2} \lfloor pN_1 \rfloor - \frac{1}{4} \frac{p_1 + p_2}{2} \left(1 - \frac{p_1 + p_2}{2}\right) (N_2 - N_1),
\end{equation*}
and define the truncated Wilcoxon test statistic as $$
T_{tW}= \frac{s^2}{\mathrm{Var}[s]}.$$ 
The statistic $s$ is very similar to $S$, except an extra term  $$\frac{1}{4} \frac{p_1 + p_2}{2} \left(1 - \frac{p_1 + p_2}{2}\right) (N_2 - N_1),$$ which is caused by the difference between the variances due to  different sample sizes. This term disappears when $N_1 = N_2$. Under the alternative hypothesis, this term is a small order term compared to the other part of $s$. Under the null hypothesis, this extra term is used to  eliminate the effect of different sample sizes, leading to an expectation close to 0. 

To calculate  $\mathrm{Var}[s]$ and to derive the asymptotic distribution of $s$,  we show that under the null that both $\mx$ and $\my$ follow the same distribution with a point mass at 0, when  $N_1 \rightarrow \infty$, $N_2 \rightarrow \infty$, we have 
\begin{equation*}\label{mean}
	\E[S] = 0, \qquad \E[s] = O(\max\{N_1, N_2\}),
\end{equation*} 
where the remainder $O(\max\{N_1, N_2\})$ is caused by the difference between $\lfloor pN_1 \rfloor$ (or $\lfloor pN_1 \rfloor$) and $p N_1$ (or $p N_2$). 
In addition, we have  
\begin{eqnarray*}
	\mathrm{Var}[S] & = & \frac{N_1^2 N_2^2}{4}\E\{(p_2 - p_1)^2\} + \frac{N_1 N_2}{12} \E(N_1 p_1^2 p_2 + N_2 p_1 p_2^2 + p_1 p_2),\\
	\mathrm{Var}[s] &=& \frac{N_1^2 N_2^2}{4} \mathrm{Var}\{(p_2 - p_1) p\}
	+ \frac{N_1 N_2}{12} \E(N_1 p_1^2 p_2 + N_2 p_1 p_2^2 + p_1 p_2),
\end{eqnarray*}
(see Supplemental Materials).
With these the results, under the null hypothesis, when $N_1 \rightarrow \infty$, $N_2 \rightarrow \infty$, and $c \leq N_1/N_2 \leq C$ for some constants $c$ and $C$, we have 
\begin{eqnarray*}
	\mathrm{Var}[S] & = & N_1 N_2 (N_1 + N_2) \theta (1 - \theta + \theta^2/3)/4 + O(N^{2.5}),\\
	\mathrm{Var}[s] &=& N_1 N_2 (N_1 + N_2) \theta^3 (1 - \theta + 1/3)/4 + O(N^{2.5}),
\end{eqnarray*}
where $\theta$ is the expected value of the proportion of non-zero observations in the data.  Hence, $\sqrt{\mathrm{Var}[s]} = O(\sqrt{N_1 N_2 \max\{N_1, N_2\}})$, and $\E[s] = O(\max\{N_1, N_2\})$ is relatively  small when $N_1$ and $N_2$ are large. Asymptotically, $s/\sqrt{\mathrm{Var}(s)}$ follows a standard normal distribution, and the truncated Wilcoxon  test statistic $T_{tW}=s^2/\mathrm{Var}[s]$  has a  $\chi^2_1$ null distribution.  

The asymptotic analysis above provides a way of approximating   $\var(s)$  using the  main term of $\var(s)$, which leads to the final   test statistic 
\[
T_{tW} = \frac{s^2}{N_1 N_2 (N_1 + N_2) (\frac{p_1+p_2}{2})^3 (4/3 - \frac{p_1+p_2}{2})/4}.
\]
This is used in our simulation and real data analysis.

\subsection{Pitman's asymptotic relative efficiency of $T_W$ and $T_{tW}$ for two-sample test}
\label{sec:wilcoxonare}
To compare the test statistics $T_W$ and $T_{tW}$, we evaluate the asymptotic property of the relative efficiency. Since we reject the null hypothesis when the statistic is  large,  the Pitman's relative efficiency is defined as
\begin{equation*}\label{ARE}
	RE(T_{tW}, T_W)=\frac{\mathrm{E}_1[T_{tW}]}{\mathrm{E}_1[T_W]} =  \frac{\mathrm{E}_1[s^2]/\mathrm{Var}[s] }{\mathrm{E}_1[S^2]/\mathrm{Var}[S]}.
\end{equation*}
Here, $\Eo$ denotes the expectation of the statistic under the  alternative hypothesis. 
A  value larger than one implies power gain using the truncated Wilcoxon test statistic  compared to the standard  Wilcoxon statistic. 

For  the two-part model \eqref{AREmodel}, 
 we need some terms to quantify  the difference between the two distributions.
Let $\theta = (\theta_1 + \theta_2)/2$, $\theta_m = \max\{\theta_1, \theta_2\}$, and $\triangle \theta = \theta_2 - \theta_1$.  
Define $\Delta_{f, g} = P(\bx < \by) + P(\bx = \by)/2-1/2$, where $\bx \sim f$ and $\by \sim g$. The term $\Delta_{f, g}$ is used to measure the effect size of the non-zero part. The following theorem provides an explicit expression for the asymptotic relative efficiency (\textsc{are}).
\begin{theorem}\label{thm:arewilcoxon}
	Under model (\ref{AREmodel}) and that $N_1 \rightarrow \infty$, $N_2 \rightarrow \infty$, and $c \leq N_1/N_2 \leq C$ holds for some constants $c$ and $C$, then the $\textsc{are}$  can be derived as 
	\begin{gather*}\label{asympA}
		\textsc{are}(T_{tW}, T_W) = \frac{1}{\theta^2} \frac{1 - \theta + \theta^2/3}{1 - \theta + 1/3} \biggl(\frac{ \theta_1 \theta_2 \Delta_{f, g} + \triangle\theta\theta_m/2}{ \theta_1 \theta_2 \Delta_{f, g} + \triangle\theta /2}\biggr)^2 + O(N^{-1/2}). \nonumber
	\end{gather*}
	Especially, we have that 
	$\mathrm{E}(s|n_1, n_2) = n_1 n_2 \Delta_{f, g}.$
\end{theorem}

To illustrate the gain in efficiency of using the truncated Wilcoxon test, we present the  $\textsc{are}(W_m, W)$ for   the following four different parameter settings to assess the effect of $\Delta_{f, g}$, $\theta$, $\triangle\theta$, and $N_1/N_2$:
\begin{itemize}
	\item[(a)] Effect of $\Delta_{f, g}$: $N_1 = 40$, and $N_2 = 50$ and $\theta_1 = 0.3$, $\theta_2 = 0.8$. Let $\Delta_{f, g} \wanjie{= -1/2 + 0.01k}$, \wanjie{where $k = 0, 1, 2, \cdots, 100$}. 
	\item[(b)] Effect of $\theta$: $N_1 = 40$,  $N_2 = 50$, and $\Delta_{f, g} = 0.1$. \wanjie{Let $\theta = 0.1 + 0.01k$, where $k = 0, 1, 2, \cdots, 80$. }
	With a given $\theta$, take $\theta_1 = \theta - 0.1$ and $\theta_2 = \theta + 0.1$.
	\item[(c)] Effect of $\triangle\theta$: $N_1 = 40$, $N_2 = 50$ and $\Delta_{f, g} = 0.1$. 
	\wanjie{Let $\triangle \theta = 0.02k$, where $k = 0, 1, 2, \cdots, 50$.}
	With a given $\triangle\theta$, take $\theta_1 = 0.5 - \triangle\theta/2$ and $\theta_2 = 0.5 + \triangle\theta/2$.
	\item[(d)] Effect of sample size: $\theta_1 = 0.3$, $\theta_2 = 0.8$, $\Delta_{f, g} = 0.1$, and $N_2 = 50$. 
	\wanjie{Choose $N_1 \in \{20, 25, 30, \cdots, 115, 120\}$. }
\end{itemize}

\begin{figure}[ht!]
	\centering
	\includegraphics[width=0.98\textwidth]{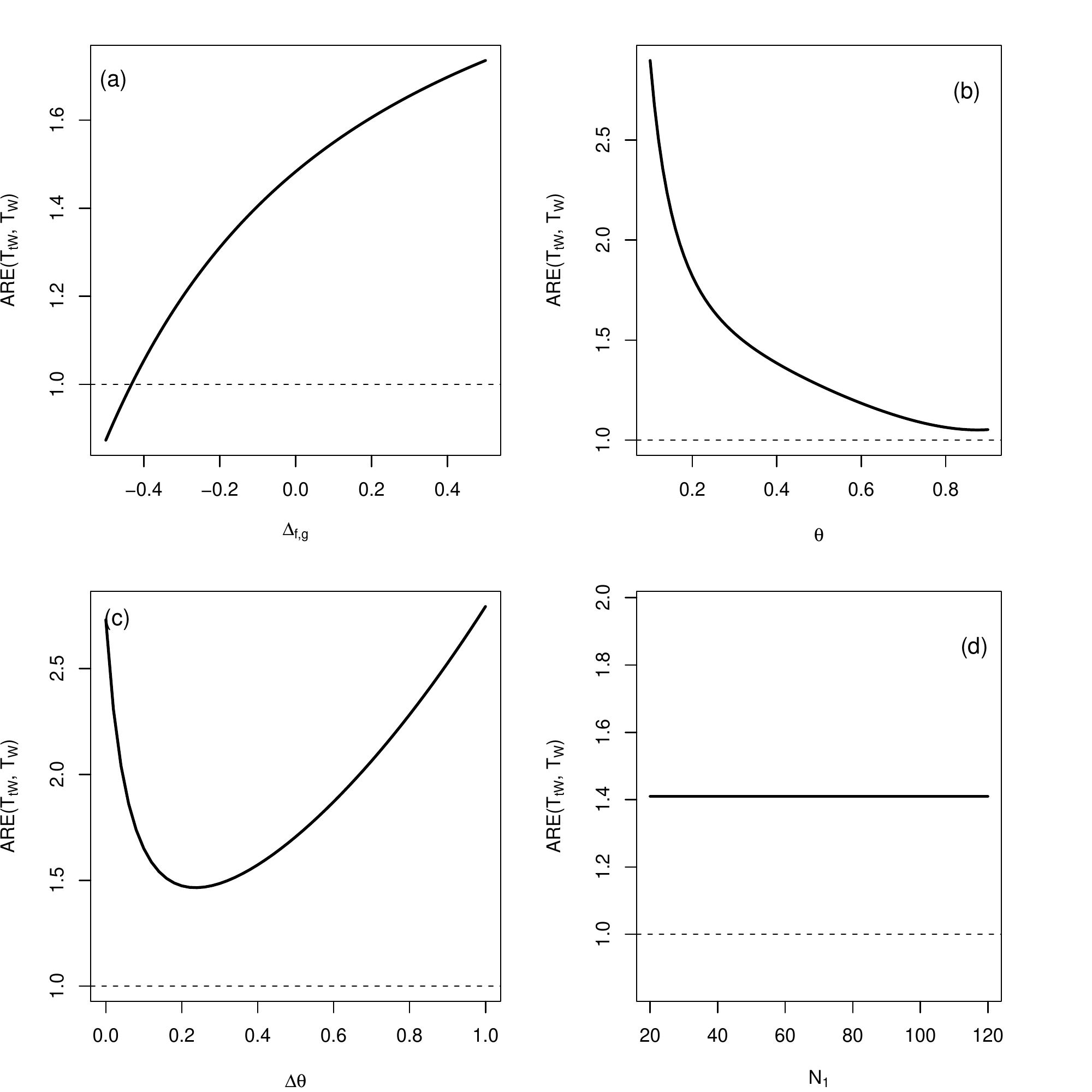}
	\caption{Asymptotic relative efficiency  $\textsc{are}(T_{tW}, T_W)$ comparing the truncated Wilcoxon rank test and the standard Wilcoxon rank test as a function of  (a) 
		$\Delta_{f, g}$, (b)  $\theta$, (c)  $\triangle\theta$,  and (d) $N_1$. \wanjie{For each plot, the horizontal dashed line represents $\textsc{are}(T_{tW}, T_W)=1$.} }\label{wilcoxon}
	
\end{figure}

Figure \ref{wilcoxon} shows the \textsc{are} for these four different settings.  We observe that   \textsc{are}s are larger than one for almost all the parameter  settings, indicating that the truncated Wilcoxon statistic  has greater power than the standard  Wilcoxon statistic. 
In settings (a) and (b), $\textsc{are}(T_{tW}, T_W)$ increases when $\Delta_{f, g}$ increases or the proportion of zeros increases. Hence, when the nonzero part are away from each other, the truncated test statistic gains more power compared to the standard Wilcoxon statistics. 
In setting  (c), the $\textsc{are}(T_{tW}, T_W)$ function is not monotone due to the cancellation of difference between non-zero part and the non-zero proportions. 
In setting  (d), obviously $\textsc{are}(T_{tW}, T_W)$ does not depend on the value of $N_1$ or $N_2$, which can be seen in the formula.  \wanjie{In Figure \ref{wilcoxon} (a), when  $\Delta_{f,g}$ has  a large negative value ($<-0.44$), $\textsc{are}(T_{tW}, T_W)$  is smaller than 1. This is the case when $\Delta_{f,g}$ and $\triangle\theta$ show opposite effects and  cancel with each other, leading to a loss of efficiency from  the  truncation of zeros. However, in real microbiome studies, this scenario is very unlikely to occur since this  would imply  that individuals who do not carry a particular bacterium have a similar risk  as the individuals who have a very high abundance of this same bacterium.  }

\wanjie{We have further verified  the theoretical $\textsc{are}$ using simulations and present the  results in Section 1 of the Supplemental Materials.  We observe that the simulated  $\textsc{are}$ is close to the theoretical $\textsc{are}$, in terms of both values and the trends as we change the parameters. 
}

\section{A truncated Kruskal-Wallis test for $K$-group comparison with equal sample sizes}\label{sec:kwequal}

\subsection{A truncated Kruskal-Wallis test}
Consider the setting where we have data from $K$ groups  $\bX_1, \bX_2, \cdots, \bX_K$, all containing non-negative independent observations  from population $1, 2, \cdots, K$, respectively. 
Each group  $\bX_i = (x_i(1), \cdots, x_i(N_i))$ contains $N_i$ $i.i.d$ observations.  
We  assume that the samples from the  $K$ groups are generated from the following two-part model, 
\begin{equation}\label{twopart}
	x_i(j) \sim  (1 - \theta_i) \delta_0 + \theta_i f_i,\quad i=1,\cdots, K.
\end{equation}
The hypothesis of interest is 
\[
H_0: \theta_1 = \cdots = \theta_K = \theta, \quad f_1 = \cdots = f_K = f \quad
vs \quad 
H_1: \mbox{not all $\theta_i$s and $f_i$s are  equal}.
\]

The Kruskal-Wallis test is a standard nonparametric test for $K$-group comparison based on the ranks.  We propose to develop a similar rank-based test that accounts for excessive zeros in the two-part model. We first consider  the case when the samples sizes from all  $K$ groups  are the same, i.e.,  $N_1 = N_2 = \cdots = N_K=N$. Let $r_{i}(j)$ be the rank of the $j$th observation in the $i$th group  among all $KN$ observations. The Kruskal-Wallis statistic can be written as  
\[
T_{KW} = \frac{12}{KN^2 (KN + 1)} \sum_{i = 1}^K S_i^2,
\]
where $S_i=\sum_{j=1}^N r_{i}(j) - (KN + 1)N/2$.  
To derive  the distribution of $T_{KW}$ under the null hypothesis, we rewrite $T_{KW}$ in terms of $Y_i$, where $Y_i = \sum_{j = 1}^{i} S_j - i S_{i + 1}$, $1 \leq i \leq K - 1$, 
\[
T_{KW} = \frac{12}{K N^2 (KN + 1)} \sum_{i = 1}^{K - 1} \frac{Y_i^2}{i(i + 1)}= \sum_{i = 1}^{K - 1} \frac{Y_i^2}{\mathrm{Var}[Y_i]}.
\]
With this transformation, $Y_i$s are asymptotically independent with each other, and under the null hypothesis,  the test statistic $T_{KW}$ is  the summation of $K - 1$ chi-square random variables with degree of freedom of 1,  and therefore  is asymptotically distributed as a $\chi^2_{K - 1}$ distribution.

In order to account for excessive  zeros in the data sets, we propose to modify the $T_{KW}$ statistic using the idea of truncation. 
Let $n_i$ denote the number of nonzero observations in group $i$, so the proportion of nonzero entries is $p_i = n_i/N$. 
Let $p = \max_{1 \leq i \leq K} p_i$. 
For each group $i$, keep the $p N$ observations with smallest ranks (or largest values), so that all the removed entries are zeros. 
Let $n = p N = \max_{1 \leq i \leq K} p_i N = \max_{1 \leq i \leq K} n_i$, then the truncated data set has $Kn$ observations in total. Rank all $Kn$ observations from smallest to largest, and then let $r_i$ denote the rank-sum of the observations in group $i$. Define 
\[
s_i = r_i - n(Kn + 1)/2, \qquad U_i = \sum_{j = 1}^{i} s_j - i s_{i+1}, \quad 1 \leq i \leq K-1,
\]
where  $s_K = - \sum_{i = 1}^{K - 1} s_i$.  The following Lemma show  that  $U_i$'s are independent.

\begin{lemma}\label{lemma:cov-m}
	Under the model that all $x_i(j)$ are independently and identically distributed, 
	\[
	\mathrm{E}[U_i] = 0, \qquad \mathrm{Cov}[U_{i_1}, U_{i_2}] = 0.
	\]
\end{lemma}

This leads to  our definition of  the truncated  Kruskal-Wallis test statistic as 
\begin{equation*}
	T_{tKW}= \sum_{i = 1}^{K - 1} \frac{U_i^2}{\mathrm{Var}[U_i]},
\end{equation*}
which has a $\chi^2$ distribution with degree of freedoms $K - 1$ under the null hypothesis. 
When $K = 2$, this statistic becomes the truncated  Wilcoxon rank-sum statistic given $N_1 = N_2$. 
 The following Lemma \ref{lemma:varu-m} provides an approximation of  the variance of $Y_i$ and $U_i$ under null hypothesis and alternative hypothesis. 
\begin{lemma}\label{lemma:varu-m}
	Consider the two-part  model (\ref{twopart}). Under the null hypothesis and suppose that $N \rightarrow \infty$, we have 
	\[
	\mathrm{Var}[Y_i] = \frac{i (i + 1) K^2 N^3 \theta}{4} \left\{\frac{\theta^2 }{3}  + (1 - \theta)\right\} + O(N^{5/2}), \quad 1 \leq i \leq K - 1;
	\]
	\begin{equation*}\label{eqn:varui}
		\mathrm{Var}[U_i] = \frac{i (i + 1) K^2 N^3 \theta^3}{4} \biggl\{\frac{1 }{3} + (1 - \theta)\biggr\} + O(N^{5/2}), \quad 1 \leq i \leq K - 1.
	\end{equation*}
	Under the alternative hypothesis, we have the upper bound of the variances, 
	\begin{equation*}
		\var_1[Y_i] \leq K^3 N^3, \quad  \var_1[U_i] \leq K^3 N^3, \qquad 1 \leq i \leq K.
	\end{equation*}
\end{lemma}

Based on this lemma, the unknown variance  $\var[U_i]$ can be approximated by $$\frac{i (i + 1) K^2 N^3 (\sum_{k=1}^K p_k/K)^3}{4} \left(\frac{4}{3}  - \sum_{k=1}^K p_k/K\right).$$ 

\subsection{Pitman's asymptotic relative efficiency  of $T_{tKW}$ and  $T_{KW}$ under a two-part model}

We evaluate the relative  efficiency of the truncated statistic $T_{tKW}$ with the standard statistic $T_{KW}$ defined by 
\[
RE(T_{tKW}, T_{KW}) = \frac{\Eo [T_{tKW}]}{\Eo [T_{KW}]} = \frac{\sum_{i = 1}^{K - 1} \mathrm{E}_1[U_i^2]/\mathrm{Var}[U_i]}{\sum_{i = 1}^{K - 1} \mathrm{E}_1[Y_i^2]/ \mathrm{Var}[Y_i]}.
\]
We assume that the $K$ groups have the same sample size  $N$ and the data are generated from the two-part model \eqref{twopart}.
To find $RE(T_{tKW}, T_{KW})$, there are four terms to calculate: the expectation of $U_i^2$ and $Y_i^2$ under the alternative hypothesis, and the variance of $U_i$ and $Y_i$ under the null hypothesis. 
To find $\Eo [U_i^2]$ and $\Eo [Y_i^2]$, we can calculate $\mathrm{Var}[U_i]$ and $\mathrm{Var}[Y_i]$, and the expectation of $U_i$ and $Y_i$ separately, where the variances are needed under either hypothesis.  Lemma \ref{lemma:varu-m} provides an approximation of  the variance of $Y_i$ and $U_i$ under null hypothesis and alternative hypothesis.

In order to calculate  the expectation of $Y_i$ and $U_i$ under alternative hypothesis, we note that the basic terms in $Y_i$ and $U_i$ are the rank-sum of the nonzero observations. 
Let $r_i^0$ denote the rank-sum  of the non-zero observations in group $i$, $1 \leq i \leq K$, and that $s_i^0 = r_i^0 - (1 + \sum_{j = 1}^K n_j)n_i/2$.  Define  $\Delta_{i, k} = P(\bx_i < \bx_k) + P(\bx_i = \bx_k)/2-1/2$,  $\bx_i \sim f_i$ and $\bx_k \sim f_k$, for $1 \leq i,k \leq K-1$, where $\Delta_{i,k}$ can be used to measure the effect sizes. 
In addition, one can easily check that
\begin{eqnarray}\label{eqn:deltadef}
	\mathrm{E}\bigl(\sum_{j = 1}^{i} s_j^0 - i s_{i + 1}^0 \big| n_1, \cdots, n_K\bigr) =  \sum_{j = 1}^i \sum_{k \neq j} n_j n_k \Delta_{j, k} - i n_{i+1}  \sum_{k \neq i+1} n_k \Delta_{i+1, k}, 
\end{eqnarray}
for $ 1 \leq i \leq K - 1$.
  For model (\ref{twopart}) and the effect sizes specified above, we have the results on $\Eo[Y_i]$ and $\Eo[U_i]$, as shown in the following Lemma. 
\begin{lemma}\label{lemma:eu}
	Under alternative hypothesis, for $1 \leq i \leq K-1$, the expectation of $Y_i$ is 
	\[
	\mathrm{E}_1[Y_i] = N^2 \biggl( \sum_{j = 1}^i \sum_{k \neq j} \theta_j \theta_k \Delta_{j, k} - i \theta_{i+1}  \sum_{k \neq i+1} \theta_k \Delta_{i+1, k} \biggr) + \frac{KN^2}{2} \biggl(i \theta_{i+1} - \sum_{j = 1}^i \theta_{j}\biggr),
	\]
	and the expectation of $U_i$ is 
	\[
	\mathrm{E}_1[U_i] = N^2 \biggl( \sum_{j = 1}^i \sum_{k \neq j} \theta_j \theta_k \Delta_{j, k} - i \theta_{i+1}  \sum_{k \neq i+1} \theta_k \Delta_{i+1, k} \biggr) + \frac{KN^2}{2} \theta_{(K)} \biggl(i \theta_{i+1} - \sum_{j = 1}^i \theta_{j}\biggr) + o(N^2).
	\]
\end{lemma}

Plugging  $\E_1[U_i]$, $\var_1[U_i]$ and $\var[U_i]$ into the definition of $\are(T_{tKW}, T_{KW})$ and using  Lemma \ref{lemma:varu-m} and Lemma \ref{lemma:eu},  we  obtain the  \textsc{are} of $T_{tKW}$ versus $T_{KW}$ given  by the following theorem. 
\begin{theorem}\label{eqn:arewilcoxon}
	Under model (\ref{twopart}) and that $N \rightarrow \infty$, then the $\textsc{are}$  can be derived as 
	\begin{equation}\label{eqn:effic}
		\textsc{are}(T_{tKW}, T_{KW}) =\frac{\sum\limits_{i = 1}^{K - 1} \frac{\bigl\{ (\sum_{j = 1}^{i} \theta_j + i \theta_{i + 1})^2 \Delta_i +  K\theta_{(K)} (i \theta_{i+1} - \sum_{j = 1}^i \theta_{j})/2 \bigr\}^2}{\theta^2 i(i + 1)(1/3 + 1 - \theta)}}
		{\sum\limits_{i = 1}^{K - 1} \frac{\bigl\{(\sum_{j = 1}^{i} \theta_j + i \theta_{i + 1})^2 \Delta_i +  K(i \theta_{i+1} - \sum_{j = 1}^i \theta_{j})/2 \bigr\}^2}{i(i + 1)(\theta^2/3 + 1 - \theta)}} + o(1).
	\end{equation}
\end{theorem}

\subsection{Asymptotic relative efficiency for zero-Beta distributions}
We consider  the case  where the nonzero functions $f_k$s are $\mathrm{Beta}$ distributions with parameters $\alpha_k$ and $\beta = 1$, in which case  the effect size  $\Delta_{j,k}$ can be calculated and   the $\textsc{are}$ can be expressed in terms of  $\alpha_k$'s and $\theta_k$'s. Since compositional data are always between 0 and 1, Beta distribution provides a reasonable parametric model for such data.

Given the Beta distribution for each sample, note that $P(\bx_i = \bx_k) = 0$ as they are continuous random variables. Hence $\Delta_{i, k} = P(\bx_i < \bx_k) - 1/2$. 
If $\bx \sim \mathrm{Beta}(\alpha, 1)$, then $\bx^\alpha \sim Unif(0, 1)$, therefore 
\begin{equation*}\label{eqn:betaunif}
	P(\bx_i < \bx_k) = P(\bx_i^{\alpha_i} < \bx_k^{\alpha_i}) = P(U < (\bx_k^{\alpha_k})^{\alpha_k/\alpha_i}), \quad 
	U \sim Unif(0, 1).
\end{equation*}
Since  $\bx_k^{\alpha_k} \sim Unif(0, 1)$, we have $(\bx_k^{\alpha_k})^{\alpha_i/\alpha_k} \sim \mathrm{Beta}(\alpha_k/\alpha_i, 1)$. Further,  for any continuous random variable $\bx$ with range $[0,1]$, there is 
\[
P(U < \bx) = \int_0^1 (1 - F_{\bx}(u)) du = \mathrm{E}[\bx]. 
\]
This leads to 
\begin{equation*}\label{eqn:betap}
	\Delta_{i, k} = P(\bx_i < \bx_k) - 1/2 = \frac{\alpha_k/\alpha_i}{\alpha_k/\alpha_i + 1} = \frac{\alpha_k}{\alpha_i + \alpha_k} - 1/2, \quad 1 \leq i, k \leq K.
\end{equation*}
When combining  with (\ref{eqn:deltadef}),  we have  
\begin{eqnarray*}\label{eqn:betar}
	&&\mathrm{E}_1[r_i^0|n_1, \cdots, n_K] = n_i\bigg\{(n_i + 1)/2 + \sum_{k \neq i} n_k \frac{\alpha_k}{\alpha_i + \alpha_k} \bigg\}, \\
	&&\mathrm{E}_1[s_i^0|n_1, \cdots, n_K] = \mathrm{E}_1[r_i^0] - \frac{1 + \sum_{k = 1}^K n_k}{2}n_i =  n_i \sum_{k \neq i} n_k \biggl(\frac{\alpha_k}{\alpha_i + \alpha_k} - 1/2\biggr).
\end{eqnarray*}
Plugging these equalities to (\ref{eqn:effic}) gives the closed-form expression for $\textsc{are}(T_{tKW}, T_{KW})$. 

\begin{figure}[ht!]
	\centering
	\includegraphics[width=0.98\textwidth]{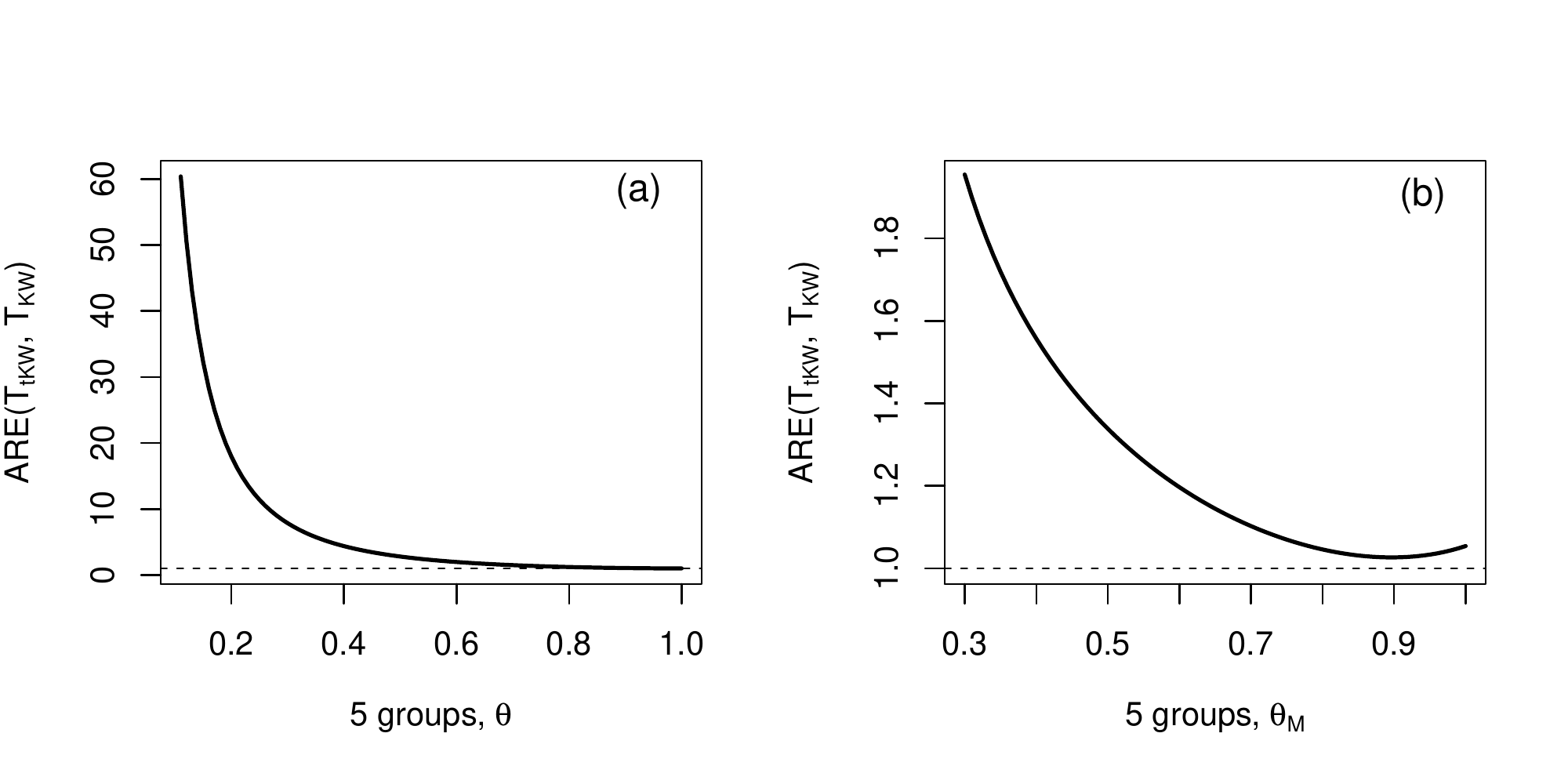}
	\caption{Asymptotic relative efficiency.  
		$\textsc{are}(T_{tKW}, T_{KW})$ as a function of  (a) $\theta$   when $\theta_1 = \cdots = \theta_K = \theta$, and (b) $\theta_{(K)}$ when not all $\theta_i$s are equal. \wanjie{For each plot, the horizontal dashed line represents $\textsc{are}(T_{tKW}, T_{KW})=1$.}
	}\label{wilcoxon1}
\end{figure}

To demonstrate the gain in efficiency, we calculate  $\textsc{are}(T_{tKW}, T_{KW})$ for five  group comparisons  ($K = 5$) in two scenarios: 
(a) $\alpha_i = i$, $1 \leq i \leq 5$. $\theta_1 = \cdots = \theta_K = \theta$, where \wanjie{$\theta = 0.1 + 0.01k$, $k = 0, 1, \cdots, 90$.}
(b) 
 $\alpha_i = i$, $1 \leq i \leq 5$. $\theta = (0.2, 0.15, 0.3, 0.1, 0.25) + d$, \wanjie{where $d = 0.01k$, $k = 0, 1, 2, \cdots, 70$.}
The results are shown in Figure \ref{wilcoxon1}. In both cases, we observe  a high asymptotic relative efficiency using the truncated Kruskal-Wallis test statistic when compared to   the original Kruskal-Wallis test. 

\section{A truncated  Kruskal-Wallis  multi-group test  with unequal  sample sizes}
\label{sec:kwdiff}
We now consider the setting where we have multiple samples $\bX_1, \bX_2, \cdots, \bX_K$, all containing non-negative independent observations from population $1, 2, \cdots, K$, respectively. 
Each group $\bX_i = (x_i(1), \cdots, x_i(N_i))$ contains $N_i$ observations sampled from the two-part model \eqref{twopart}. 
We consider the case that $N_1$, $N_2$, $\cdots$, $N_K$ are not necessarily equal.  

Consider the standard Kruskal-Wallis statistic first. Let $r_{i}(j)$ be the rank of the $j$th observation in the $i$th group among all the observations. Define $S_i=\sum_{j=1}^{N_i} r_{i}(j) - (\sum_{i=1}^K N_i + 1)N_i/2$  and 
$Y_i = \sum_{j = 1}^{i} (N_{i+1} S_j - N_j S_{i+1})$, $1 \leq i \leq K - 1$,  where $Y_i$ is also related to the sample size. The Kruskal-Wallis statistic is defined as 
\[
T_{KW} = \frac{12}{\sum_{j = 1}^K N_j + 1}  \sum_{i = 1}^{K - 1} \frac{Y_i^2}{N_{i+1} (\sum_{j = 1}^{i} N_j) (\sum_{j = 1}^{i+1} N_j) (\sum_{j = 1}^{K} N_j)}, \quad 1 \leq i \leq K-1.
\]
Each term follows a $\chi^2_1$ distribution asymptotically and is independent with each other, so the statistic follows $\chi^2_{K-1}$ distribution under the  null hypothesis.

To account for zeros, for group $i$, there are $n_i$ non-zero elements, and the corresponding non-zero ratio is $p_i = n_i/N_i$. Let $p = \max\limits_{1 \leq i \leq K} p_i$, and keep the $\lfloor p N_i \rfloor$ largest observations only for group $i$ as the truncated data. All the removed entries are zeros. 
Let $r_i$ be the rank-sum for the truncated samples in group $i$, and 
$$s_i = r_i - \frac{\lfloor p\sum_{i = 1}^K N_i \rfloor + 1}{2} \lfloor p N_i \rfloor.$$ 
We define  the statistic $U_i$ as the counterpart of $Y_i$ in the  standard Kruskal-Wallis statistic, 
\begin{equation*}\label{eqn:anova}
	U_i = \sum_{j = 1}^{i} \bigl(N_{i+1} s_j - N_j s_{i + 1}\bigr), \qquad 	1 \leq i \leq K-1.
\end{equation*}
Then, a natural test statistic based on truncation is 
\begin{equation}\label{def:kwdiff}
	T_{tKW} = \sum_{i = 1}^{K - 1} \frac{U_i^2}{\mathrm{Var}[U_i]},
\end{equation}
where $\mathrm{Var}[U_i]$ is the variance of $U_i$ under the null. This is calculated by noting that 
$\mathrm{Var}[U_i] =  \mathrm{Var}\{\mathrm{E}(U_i|p_1, \cdots, p_K)\} + \mathrm{E}\{\mathrm{Var}(U_i|p_1, \cdots, p_K)\}$. 
Lemma \ref{lemma:diffvaru} in Supplemental Materials shows that 
	\begin{eqnarray*}
		\mathrm{Var}[U_i|p_1, \cdots, p_K] &=&
\frac{1}{12}\mathrm{E}\biggl[ (\sum\limits_{k = 1}^{K} n_k + 1) \biggl\{\sum\limits_{k = 1}^{K} n_k [N_{i+1}^2 \sum\limits_{j = 1}^{i} n_j + n_{i+1} (\sum\limits_{j = 1}^{i} N_j)^2]  \\
 && -(n_{i+1}\sum\limits_{j = 1}^{i} N_j - N_{i+1}\sum\limits_{j = 1}^{i} n_j)^2 \biggr\} \biggr],\\
 \mathrm{E}[\mathrm{Var}[U_i|p_1, \cdots, p_K]] & = & \frac{\theta^2}{12} N_{i+1} (\sum_{j=1}^i N_j)(\sum_{j=1}^{i+1} N_j)(\sum_{j=1}^K N_j) \bigl[(\sum_{j=1}^K N_j)\theta + 3 - 2\theta\bigr].
 \end{eqnarray*}
However, when the sample sizes are not equal,  it is difficult to evaluate  $\mathrm{Var}[U_i|p_1, \cdots, p_K]$. Instead we  approximate this  under the null hypothesis by  assuming that the non-zero probability $\theta$ is the average of the empirical non-zero probability among all samples and by simulations since   $\mathrm{Var}[U_i|p_1, \cdots, p_K]$ only depends on the non-zero proportions, but not on $f$.

Finally, in order to prove   that $T_{tKW}$ has an asymptotic distribution of  $\chi^2_{K-1}$, we show that $(\mathrm{E}[U_i])^2$ is  much smaller than $\mathrm{Var}[U_i]$ so that $\mathrm{E}[U_i]/\sqrt{\mathrm{Var}[U_i]}$ is approximately 0, and the correlation between either two terms is asymptotically 0. Combining  the upper bound of $\mathrm{E}[U_i]$ given in Lemma \ref{lemma:expbound} and the lower bound for the variance term given in Lemma \ref{lemma:diffvaru} in the Supplemental Material, we show that  $\mathrm{E}[U_i]/\sqrt{\mathrm{Var}[U_i]} \rightarrow 0$. Using the upper bound for the covariance in Lemma \ref{lemma:diffcov},  we see that when $\max_{1 \leq i \leq K} N_i/N_{(1)}^2 \rightarrow 0$ and $N_{(1)} \rightarrow \infty$, we have 
\[
|\mathrm{Cor}[U_i, U_j]| = |\frac{\mathrm{Cov}[U_i, U_j]}{\sqrt{\mathrm{Var}[U_i] \mathrm{Var}[U_j]}}| \leq C\sqrt{(\sum_{k = 1}^K \frac{1}{N_k})} \leq \frac{C \sqrt{K}}{\sqrt{N_{(1)}}} \rightarrow 0, 1\leq i, j \leq K-1.
\]
Therefore, as long as the sample sizes are  on the same order and go to infinity, $(U_i/\mathrm{Var}[U_i], i=1,\cdots,K)$   has an asymptotic multivariate normal distribution with  mean zero   and identity covariance matrix. 
We leave the details of these lemmas in the Supplemental Materials.

\section{Simulations}\label{sec:simu}
To further verify the gain in efficiency in using the proposed truncated rank-based tests, we present simulation studies to evaluate the proposed tests and  to compare with the  standard  Wilcoxon rank-sum and Kruskal-Wallis test statistics. 
In each of the simulation setups, we perform the following steps.
\begin{enumerate}
	\item Given parameters $(K, N, \theta, \alpha, \beta, s)$, where $N$, $\theta$, $\alpha$ and $\beta$ are all $K \times 1$ arrays,  generate $\bx$  from the distribution
	\[
	x_{i}(j) \stackrel{i.i.d}{\sim} (1 - \theta_i) \delta_0 + \theta_i \mathrm{Beta}(\alpha_i, \beta_i), \quad 1 \leq i \leq K, 1 \leq j \leq N_i.
	\]
	\item Calculate the p-value from usual rank tests, including the Wilcoxon rank-sum test for two-sample test, and Kruskal-Wallis test for more groups, and our proposed tests. 
	\item Repeat steps 1-2 for $M$ times, and calculate the power or type I error  for a given  significance level $\alpha$. 
\end{enumerate}

\subsection{Simulation  1 - evaluation of Type I error}

We first evaluate the type I errors of various test statistics proposed in this paper  and compare them to   Wilcoxon rank-sum test and Kruskal-Wallis test.  To simulate data from the null distribution, for each group $i$, we simulate data from a two-part model, 
$$x_{i}(j) \stackrel{i.i.d}{\sim} (1-\theta) \delta_0 + \theta \mathrm{Beta}(a, b),$$
where  $a = b = 2$, $\theta = 0.5$.  We consider 
three different scenarios :  
\begin{enumerate}
	\item[(a)] $K = 2$, sample sizes = $(0.65, 1)\times N$;
	\item[(b)]  $K = 3$, equal sample sizes $N$; 
	\item[(c)] $K = 3$, sample sizes= $(0.7, 1, 1.5)\times N$. 
\end{enumerate}For each scenario,  take $N = \{30, 60, 100, 300, 600, 900\}$, and simulate $100,000$ test statistics for each choice of $N$. 
The empirical Type I errors are  summarized in Table \ref{table:exp1} for significance level  $\alpha \in\{0.05, 0.01, 0.001 \}$.  In general, we observe that the proposed tests have correct Type I errors, especially when the sample sizes are large. \wanjie{For small sample sizes, the proposed tests are slightly anti-conservative, indicating the asymptotic approximation of the test statistics may require relatively large sample sizes. In practice, when the sample sizes are small, one can obtain more accurate   $p$-values based on permutations. }

\begin{table}[ht!]
	\caption{Comparison of empirical Type I errors for  two-group test with unequal sample sizes and three-group tests with equal and unequal sample sizes. $T_W$: Wilcoxon rank-sum test; $T_{tW}$: truncated Wilcoxon rank-sum test; $T_{KW}$: Kruskal-Wallis test; $T_{tKW}$: truncated Kruskal-Wallis test.}
	\begin{center}
		\begin{tabular}{llccccc}
			\hline
			\multicolumn{2}{c}{}&\multicolumn{5}{c}{Two-group - unequal sample sizes}\\
			$\alpha$-level&  & (20, 30)  & (39, 60)  & (65, 100) & (195, 300) & (390, 600) \\
			\cline{3-7}
			\multirow{2}{*}{$0.05$} 
			& $T_W$  &.049  &.049  &.050  &.049 	&.049 	\\
			& $T_{tW}$  &.054  &.051  &.051  &.051 	& .050  \\
			\multirow{2}{*}{$0.01$} 
			& $T_W$  &.009  &.010  &.010  &.010 	&.010 	\\
			& $T_{tW}$  &.014  &.013  &.012  &.011  &.011 	\\
			\multirow{2}{*}{$0.001$} 
			& $T_W$   & $6.8\times 10^{-4}$   &$9.3\times 10^{-4}$  &$9.1\times 10^{-4}$  &$1.2\times 10^{-3}$  & $9.9\times 10^{-4}$    \\
			& $T_{tW}$  &$2.2 \times 10^{-3}$    &$1.9 \times 10^{-3}$ & $1.8\times 10^{-3}$  &$1.6\times 10^{-3}$  &$1.2\times 10^{-3}$ \\
			\hline
			\multicolumn{2}{c}{}&\multicolumn{5}{c}{Three-group - equal sample sizes}\\
			&& 30  & 60  & 100 & 300 & 600 \\
			\cline{3-7}
			\multirow{2}{*}{$0.05$} 
			& $T_{KW}$  &.049  &.048  &.049  &.050 	&.050 \\
			& $KM_m$  &.056  &.053  &.053  &.052 	& .051  \\
			\multirow{2}{*}{$0.01$} 
			& $T_{KW}$  &.009  &.009  &.010  &.010 	&.010 	\\
			& $T_{tKW}$  &.015  &.013  &.013  &.012  &.011 \\
			\multirow{2}{*}{$0.001$} 
			& $T_{KW}$   & $6.7\times 10^{-4}$   &$8.7\times 10^{-4}$  &$1.1\times 10^{-3}$  &$9.9\times 10^{-4}$  & $1.0\times 10^{-3}$     \\
			& $T_{tKW}$  &$2.5 \times 10^{-3}$    &$2.0 \times 10^{-3}$ & $1.9\times 10^{-3}$  &$1.6\times 10^{-3}$  &$1.2\times 10^{-3}$ \\
			\multicolumn{2}{c}{}&\multicolumn{5}{c}{Three-group  - unequal sample sizes}			\\
			& & (21, 30, 45)  & (42, 60, 90)  & (70, 100, 150) & (210, 300, 450) & (420, 600, 900)\\
			\cline{3-7}
			\multirow{2}{*}{$0.05$} 
			& $T_{KW}$  &.047  &.050  &.049  &.049 	&.049\\
			& $T_{tKW}$  &.055  &.056  &.054  &.051 	& .051\\
			\multirow{2}{*}{$0.01$} 
			& $T_{KW}$  &.009  &.010  &.010  &.010 	&.010\\
			& $T_{tKW}$  &.014  &.014  &.013  &.012  &.011\\
			\multirow{2}{*}{$0.001$} 
			& $T_{KW}$   & $6.4\times 10^{-4}$   &$8.1\times 10^{-4}$  &$1.0\times 10^{-3}$  &$8.9\times 10^{-4}$  & $8.6\times 10^{-4}$\\
			& $T_{tKW}$  &$2.5 \times 10^{-3}$    &$2.1 \times 10^{-3}$ & $1.9\times 10^{-3}$  &$1.4\times 10^{-3}$  &$1.3\times 10^{-3}$ \\
			\hline
		\end{tabular}
		\label{table:exp1}
	\end{center}
\end{table}

\subsection{Simulation 2 - power comparisons}

\begin{figure}[!ht]
	\centering
	\includegraphics[width=0.98\textwidth,height=0.82\textheight]{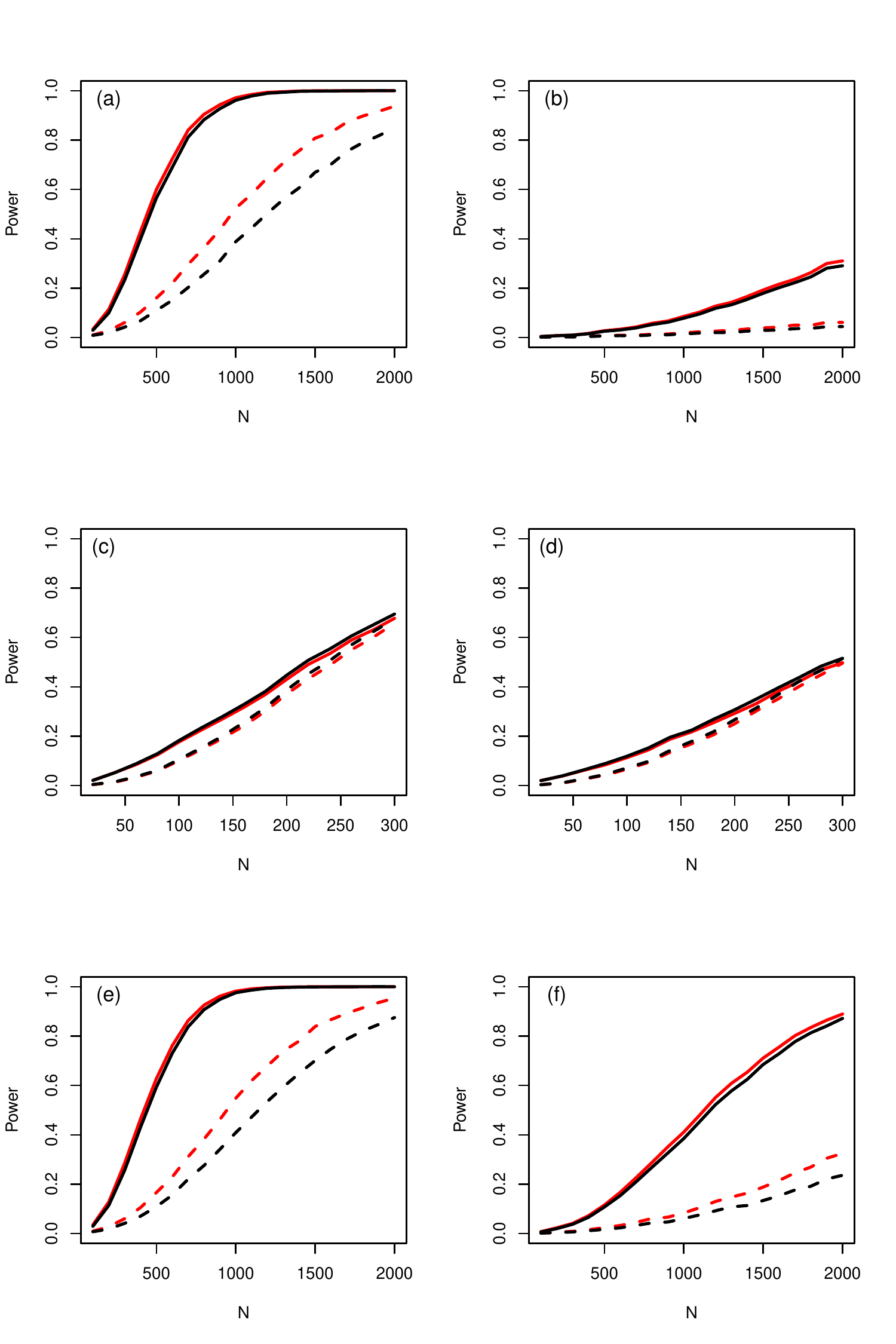}
\vspace{-0.2in}
	\caption{Power curves for the truncated Kruskal-Wallis (solid line) and the Kruskal-Wallis  test (dashed line) as a function of the sample size $N$.  (a)-(b): a two-part model with $\theta = (0.5, 0.5, 0.5)$,
		$\alpha = (1.5, 2, 2.5)$, $\beta = (2, 2, 2)$ and (b) $\alpha = (1.7, 2, 2.3)$, $\beta = (2, 2, 2)$.
		(c)-(d):  a two-part model with  $(\alpha, \beta) = \{(1/2, 1/2, 1/2), (2, 2, 2)\}$, $\theta \in \{(0.10, 0.15, 0.22), (0.15, 0.20, 0.27)\}$. 
		(e)-(f):  a two-part model with $\theta=  (.4, .5, .6)$ and  $\alpha \in \{(1.5, 2, 2.5), (1.7, 2, 2.3)\}$, and $\beta =(2, 2, 2)$. \wanjie{For each plot, red lines represent the power curves with reduced sequencing depths.} }\label{exp3}
\end{figure}

We next evaluate the power of the proposed tests. We consider three different models with $K=3$ groups and repeat the simulation $M=10000$ times.  For the first model,
we assume that the proportions of zeros are the same across different groups and examine how the distribution of the non-zero observation   affects the power of the proposed tests. 
We set the  $\theta =  (0.5, 0.5, 0.5)$ and equal sample sizes  for the three   groups,  
\wanjie{chosen from the set $\{100, 200, 300, \cdots, 1900, 2000\}$.}
 We set   $\alpha \in \{(1.5, 2, 2.5), (1.7, 2, 2.3)\}$, and $\beta =(2, 2, 2)$ and calculate the power function for different sample sizes $N$. The resulting  power curves  are shown in  the top row of Figure \ref{exp3}.  We observe a substantial gain in power from the truncated Kruskal-Wallis test.

For the second model,  we study how  different proportions of zeros  affect the test  power as the sample size  $N$ changes. We set  $(\alpha, \beta) = \{(1/2, 1/2, 1/2), (2, 2, 2)\}$, which assumes that  the nonzero distributions are the same across different groups. We again assume that the sample sizes are the same for all three groups,  \wanjie{chosen from the set $\{20, 40, 60, \cdots, 280, 300\}$.}
We set  $\theta \in \{(0.1, 0.15, 0.22), (0.15, 0.20, 0.27)\}$, and calculate the power function for different sample sizes $N$.
The resulting power curves are shown in  the middle row of Figure \ref{exp3}. Our test has some improvement over the original test. However, in this case, the improvement is not as large as when $f_i$s are different.

For the last model,  we evaluate the proposed tests when the sample sizes are different in different groups.  We set $\theta= (0.4, 0.5, 0.6)$ and  the sample size as $N \times (0.8, 1, 1.5)$, where \wanjie{$N \in \{100, 200, 300, \cdots, 1900, 2000\}$.}
We  choose $\alpha \in \{(1.5, 2, 2.5), (1.7, 2, 2.3)\}$ and $\beta = (2, 2, 2)$, and calculate the power function for  different sample sizes $N$. 
The power curves are shown in the bottom row of Figure \ref{exp3}, showing that the truncated  Kruskal-Wallis test has much higher power than the Kruskal-Wallis test when there are ties.

\wanjie{We finally examine the sensitivity of the proposed tests to reduced sequencing   depths.  One effect of having low sequencing  depth is that some rare bacteria might not be sequenced, which results in zero counts for bacteria with low but non-zero true relative abundance in the final microbial composition. Specifically,   after we generate the compositional data  for each population,  to minic lower sequencing  depth, we set the abundance of the bacteria with small relative abundance to zero with a probability of 0.5, resulting about 5\% of non-zeros proportions being set to zero.   The final data sets include more zeros due to reduced sequencing depths. We obtain the empirical power again based on the final data sets.  The new power curves are shown in 
	Figure \ref{exp3} (red lines). The proposed test still achieve higher power than the standard tests (dashed line).
	Overall, we see that the proposed tests are not too sensitive to read depths from sequencing. 
}

\section{Identifying the Crohn's disease-associated bacterial genera and the effects of treatment}
\label{sec:data}
Crohn's disease, a chronic inflammatory bowel disease, is characterized by altered composition of the gut microbiota or dysbiosis. The etiology and clinical significance of the dysbiosis is unknown.  In a recent study at the University of Pennsylvania,   the composition of the gut microbiota among a cohort of 85  children with Crohn's disease who were initiating therapy with either a defined formula diet  ($n$=33) or an anti-tumor necrosis factor $\alpha$ (anti-TNF) drug ($n$=52) was examined in order to better understand the cause of the dysbiosis in Crohn's disease.  Fecal samples were collected at baseline, 1, 4, and 8 weeks and DNA content characterized by shot-gun genomic sequencing with $>$0.5 tetra-bytes  of total sequences \citep{IBD}.   The gut microbiota data  of 26 normal children with no known gastrointestinal disorders were similarly collected.   The MetaPhlAn program \citep{segata2012metagenomic} was applied to  first obtain the relative proportions of the bacterial genera  for each of the normal and Crohn's disease samples. The number of bacterial genera  called by the MetaPhlAn program varied  from sample to sample.  Altogether,  60 bacterial genera  were observed in both healthy samples and the Crohn's disease samples.  Figure \ref{fig:heatmap} shows the relative abundances of the 60 genera in all the samples, where large proportions of zeros are observed for many of the genera.

\ignore{
\begin{figure}[!ht]
	\centering
	\includegraphics[width=0.98\textwidth, height=0.35\textheight]{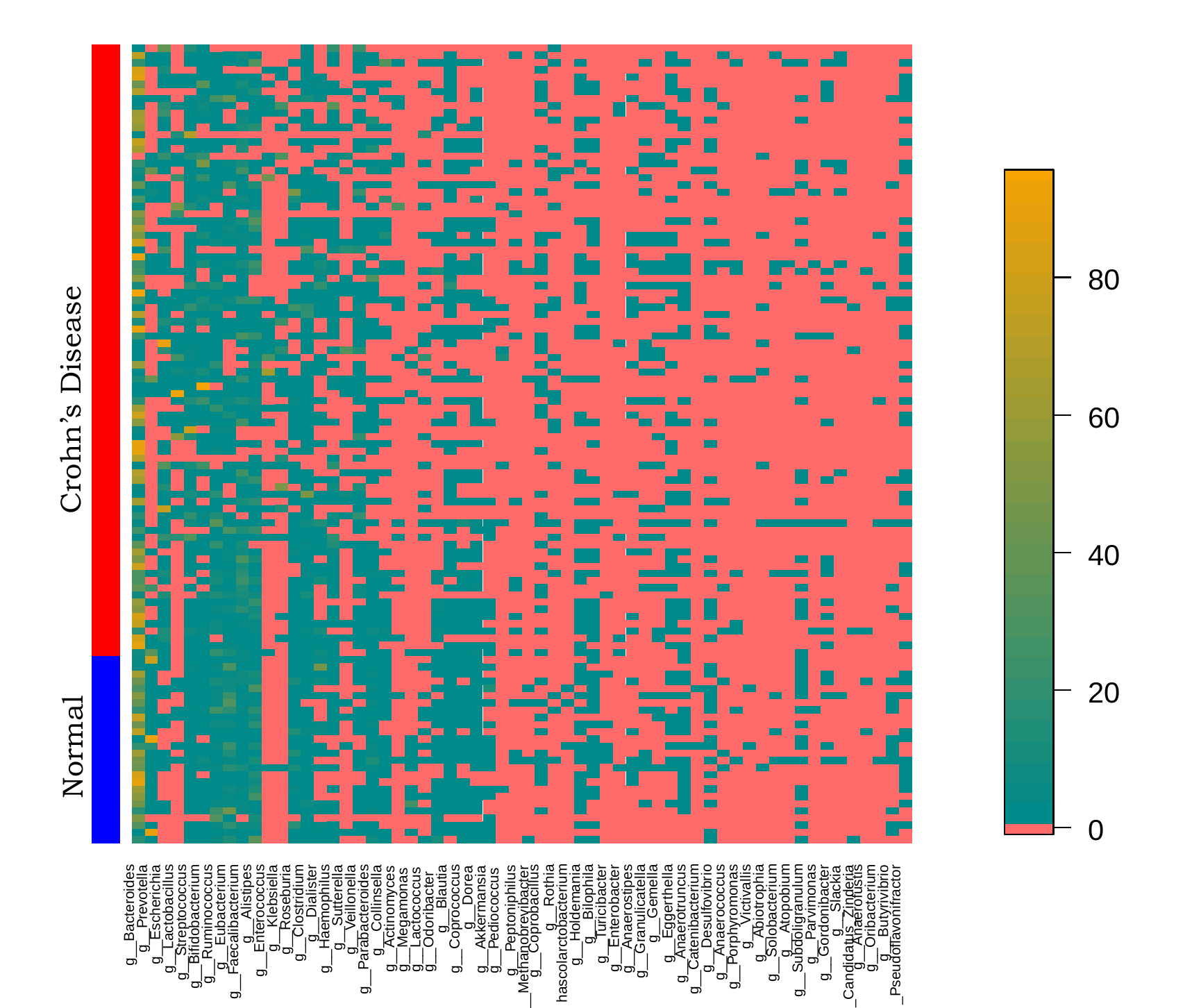}
	\caption{Heatmap of zeros   and the relative abundances for 26 normal samples  and 85 samples in Crohn's disease group for each of the 60 bacterial genera. } 
	\label{fig:heatmap}
\end{figure}
}

\subsection{Comparison of gut microbiome between normal and Crohn's disease patients}

\begin{figure}[!ht]
	\centering
	\includegraphics[width = 0.99\textwidth, height=0.39\textheight]{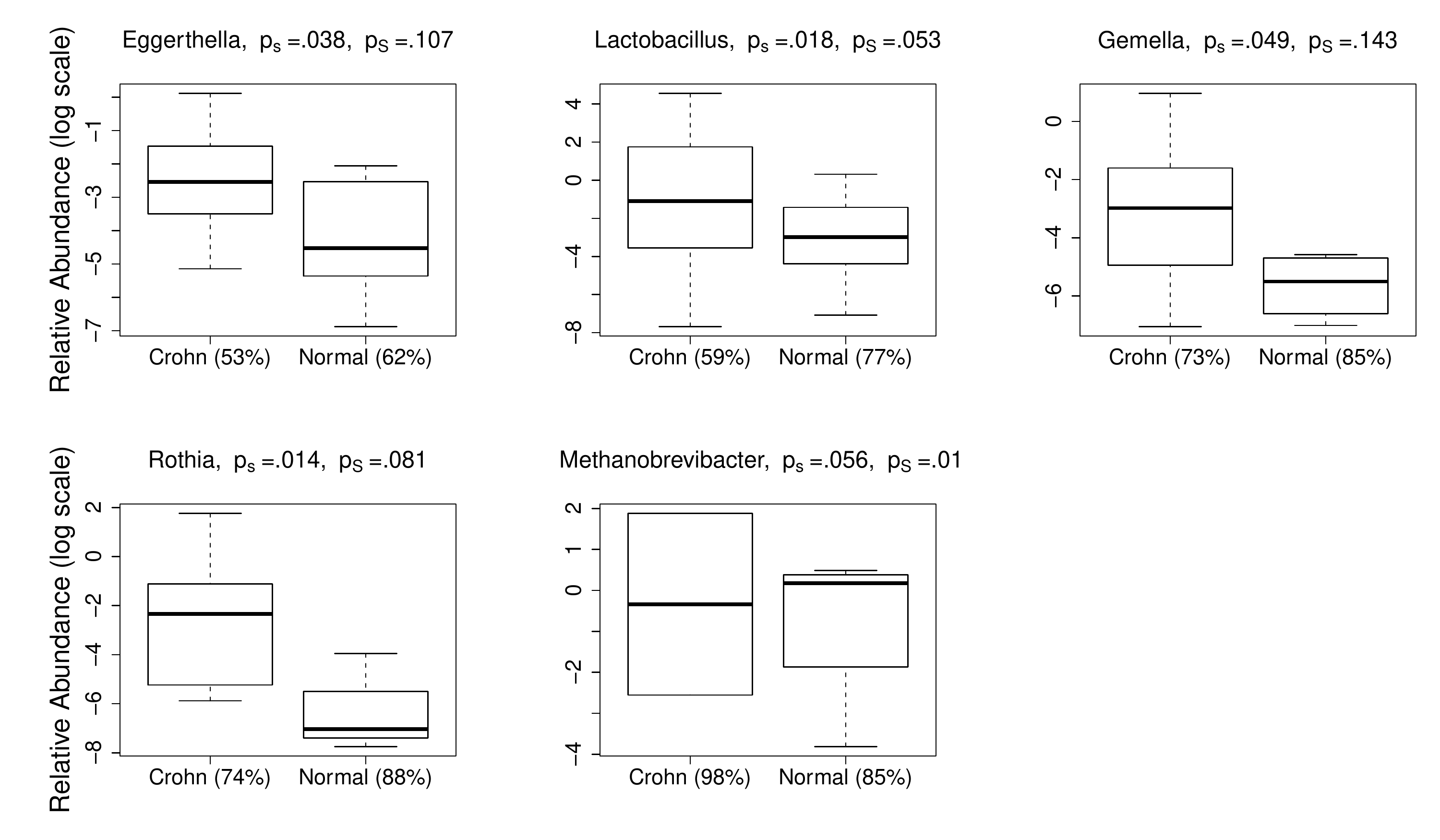}
	\caption{The box plots of the non-zero relative abundances \wanjie{(on  log scale)} with proportions of zeros in $x$-axis labels for the four bacterial genera  that were identified by the truncated Wilcoxon rank-sum test only for a nominal $\alpha$ level of 0.05, where $p$-values from $T_W$ and $T_{tW}$ tests are shown as $p_S$ and $p_s$. The genus  {\it Methanobrevibacterm}  was identified only by the standard Wilcoxon rank-sum test.  }
	\label{plot1}
\end{figure}

We are  interested in identifying the bacterial genera that show different distributions between healthy  and Crohn's disease children. At an $\alpha$ level of 0.05, the truncated Wilcoxon rank-sum test identified 23 genera that showed different distributions between healthy children  and patients with Crohn's disease, while the standard Wilcoxon test identified 20, among these, 19 were identified by both methods. At FDR of 10\%,  the truncated Wilcoxon rank-sum test identified 21 genera and the standard  Wilcoxon test identified the same 20 genera.  Four genera, including {\it Eggerthella, Lactobacillus, Gemella}, and {\it Rothia} were identified only by the truncated Wilcoxon rank-sum test.   Figure \ref{plot1} shows the the proportions of zeros and boxplots of these four genera that  clearly show difference in abundances in healthy  and Crohn's disease children, both in terms of proportion of zeros and the median of non-zero abundances.  All four genera had higher abundances in Crohn's patients than the healthy  controls.  Among these genera, {\it Eggerthella, Lactobacillus}, both are anaerobic, non-sporulating, Gram-positive bacilli, have been reported to be associated with clinically significant bacteraemia 
and  Crohn's disease \citep{Crohn1}.  In contrast,  for genus  {\it Methanobrevibacter}, the standard Wilcoxon test had a sightly smaller $p$-value.  However, 98\% of the disease individuals did not carry this genus. \wanjie{In this case, the truncation may lead to a slightly reduced statistical significance. However, this can also be due to   random error or the asymptotic approximation error due to relatively small sample sizes. To verify this, we  performed  100,000 permutations of group labels  and obtained a p-value of 0.016 and 0.017, for the truncated test and the standard Wilcoxon test, respectively. }

\subsection{Comparison of gut microbiome across time after treatment}

\begin{figure}[!ht]
	\centering
	\includegraphics[width = 0.99\textwidth, height=0.33\textheight]{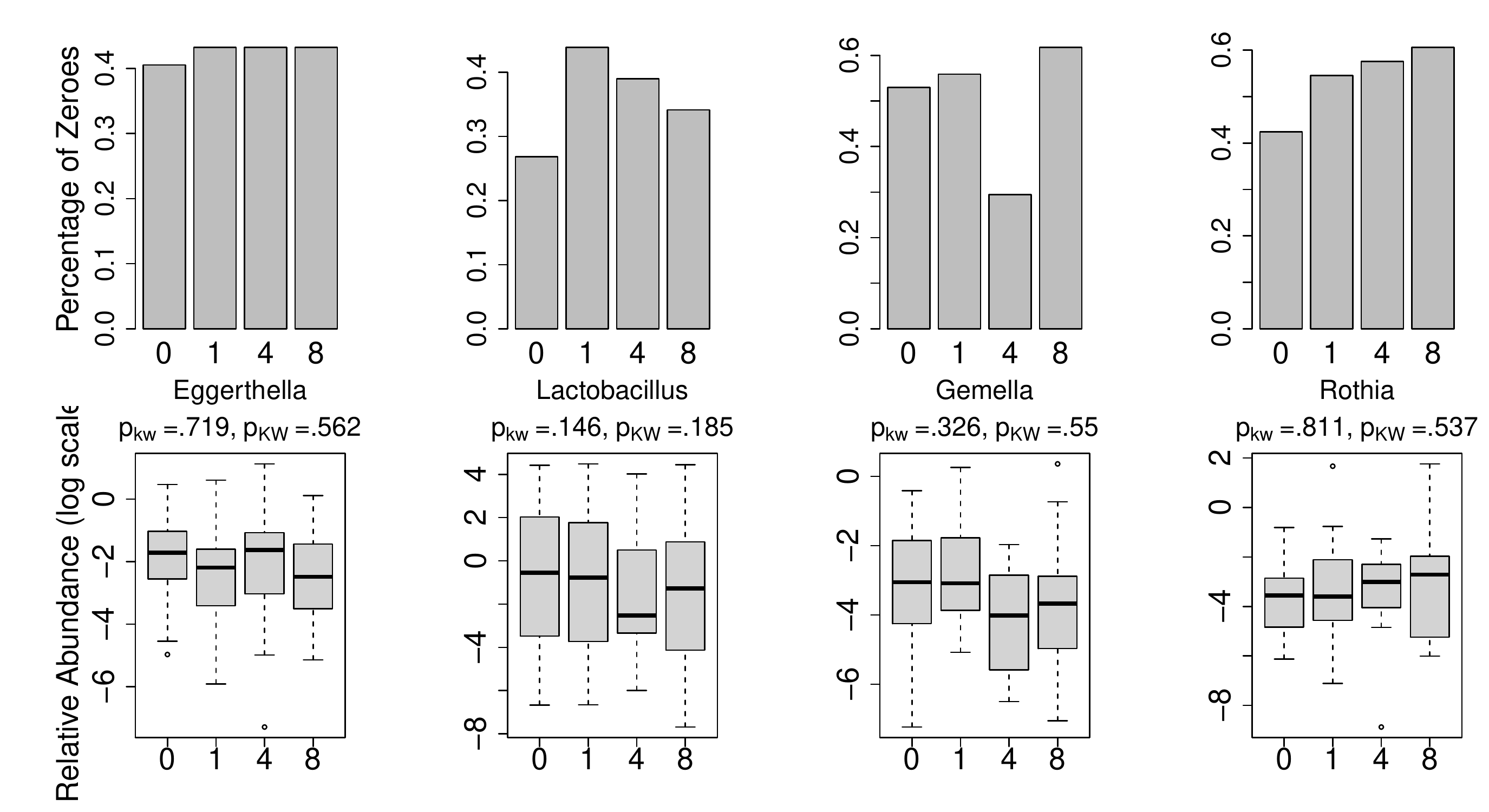}
	\caption{Bar plots of proportions of zeros (top row) and box plots of the relative non-zero abundances (bottom row\wanjie{, on log scale}) over four time points during anti-TNF treatment for four  bacteria genera that were identified only by the truncated Kruskal-Wallis  test. }
	\label{plot2}
\end{figure}

We next aim to identify the bacterial genera that show changes of relative abundances across the four time points during the anti-TNF treatment.  We applied our truncated Kruskal-Wallis test statistic with within-subject permutations (100,000 permutations) and identified 7  genera with changes of abundances during the treatment with $p<0.05$. As a comparison, the Friedman's test identified only three.  Figure \ref{plot2} shows the boxplots of the abundances of the genera across 4 time points for the four  genera identified only by our truncated Kruskal-Wallis test, including {\it Bacteroides, Roseburia, Eubacterium} and {\it Bilophila}.   Among these, {\it Bacteroides, Eubacterium} and {\it Bilophila}  showed increased abundances at 8 weeks after the anti-TNF treatment.  Interestingly, all three genera have been shown to have reduced abundance in Crohn's disease patients \citep{Crohn2}.   
Reduction of  {\it Roseburia}, a  well-known butyrate-producing bacterium  of the Firmicutes phylum, has been consistently demonstrated to be associated with Crohn's disease \citep{Crohn3}.  There has been  evidence that the gut bacteria in patients with inflammatory bowel disease do not make butyrate, and that they have low levels of the fatty acid in their gut \citep{Sartor}.  The decreased abundance of {\it Bilophila}, may translate into a reduction of commensal bacteria-mediated, anti-inflammatory activities in the mucosa, which are relevant to the pathophysiology of Crohn's disease. 
The result shows the effect of anti-TNF treatment in increasing the relative abundances of {\it Roseburia, Bacteroides} and {\it Bilophila},   therefore potentially increasing the level of fatty acid butyrate and anti-inflammatory activities.    This partially explains that 50\% of the patients showed clinical improvement, as reflected by reduction of fecal calprotection below 250 mcg/g \citep{IBD}.

\section{Discussion}\label{sec:diss}
Motivated by comparing the distributions  of taxa composition in different groups in microbiome studies, we have developed  several extensions of the popular rank-based tests to account for clumps of zeros, including truncated rank-based Wilcoxon and Kruskal-Wallis tests for two- or multiple-group  comparisons.  These tests are rank-based and nonparametric and are easy to implement.  By using within-sample permutations, such tests can also be applied to paired samples or repeated measurements analysis. We have shown that the proposed tests have better power than the standard rank-based tests, both by asymptotic relative efficiency analysis  and by simulations.  We observed a large gain in power when the proportions of zeros in the data sets are high as compared to the standard rank-based tests due the high number of tied ranks. \cite{Hallstrom}  showed that the truncated Wilcoxon rank-sum test has equal or better power than the test of  two-degree of freedoms  based on two-part model when the sample sizes are equal.   \wanjie{Since our two-sample truncated test is an extension of Hallstrom's test to unequal sample sizes, we  expect that our proposed tests should have similar or better power than the two-part tests that combine binomial test and nonparametric test for the continuous part.} Although such two-part tests can be extended to two-part model for multiple group comparisons, it has not been studied in literature.  It would be interesting to compare the performance of our proposed truncated Kruskal-Wallis tests with other  tests based on the two-part models. 

We have demonstrated the applications of the proposed tests in an  analysis of real metagenomic data sets. Our results have shown that the truncated rank-based tests are effective and  identify more bacterial genera that are associated with the clinical phenotypes and treatment.   As observed in other studies \citep{Wagner}, tests that account for clumps of zeros  can be more powerful in testing abundance difference  in microbiome studies. We have demonstrated that some well-known Crohn's disease-associated bacterial genera can be missed by using the standard rank-based tests without adjusting for excessive  zeros.  We  expect to see more  applications of the proposed  tests in microbiome studies.

\wanjie{The proposed tests have several limitations. First, since these tests are rank-based and nonparametric, they cannot directly account for covariate effects in observational studies. 
 If there is no severe covariate imbalance, we can perform quantile-stratification based on covariates and apply the proposed tests to each strata and then combine the results using e.g., Fisher's combination of $p$-values. This requires large sample sizes. Second,  since the $p$-values of our proposed rank-based tests are calculated based on the asymptotic distributions of the test statistics, which require relatively large sample sizes, they may not be accurate enough when the sample sizes are small. As shown in  our Table 1, as the sample sizes increase, the Type I error gets closer to the nominal level.  For small sample sizes, we would suggest that the users apply the proposed tests and obtain the $p$-values based the asymptotic distributions. Once the taxa are identified, one can run a permutation test to further confirm the results. This will save time to run permutation tests for all the taxa. 
}
\section*{Acknowledgment}
This research was supported by NIH grants GM129781 and GM123056. We thank Dr. Kafadar, the AE and two reviewers for many helpful comments and suggestions.   

\section*{SUPPLEMENTARY MATERIALS} 
The online Supplemental Materials include proofs of Theorem 1, Theorem 2 and all the lemmas. It also includes detailed derivations of the proposed test statistics and simulations to verify the theoretical asymptotic relative efficiency of the proposed tests. 
{An R repository of the proposed method and all analyses performed is available at 
	\texttt{https://github.com/hongzhe88/Truncated-Rank-based-Tests}.

\bibliographystyle{imsart-nameyear}
\bibliography{refs}

\newpage
\title{Supplementary Materials}
\appendix

\section{Simulations on ARE$(T_{tKW}, T_{KW})$}\label{sec:are}
In Theorem \ref{thm:arewilcoxon}, we  derived the theoretical results of $\are(T_{tKW}, T_{KW})$ and  presented how it relates to different  parameters in Figure \ref{wilcoxon}. 
In this section, we present simulation results to further verify the theoretial results  presented in Figure \ref{wilcoxon}. 

For each of the 4 settings, we calculated the \textsc{are} by empirical means and variances of the test statistics under null and alternatives based on 10,000 simulations.  The four settings are almost the same with that in Figure \ref{wilcoxon}, except some extreme values that are hard to realize in numerical studies. 
\begin{itemize}
	\item[(a)] Effect of $\Delta_{f, g}$: $N_1 = 40$, and $N_2 = 50$ and $\theta_1 = 0.3$, $\theta_2 = 0.8$. Let $\Delta_{f, g} {= -0.5 + 0.01k}$, {where $k = 3, 4, 5, \cdots, 98$}. 
	For each choice $\delta = \Delta_{f,g}$, we take $f \sim Beta(3, 3)$, and $g \sim Beta(\alpha, \beta)$, where $1 < \alpha < 10$ and $1 < \beta < 10$ are chosen so that $|\Delta_{f,g} - \delta| < 0.05$.
	\item[(b)] Effect of $\theta$: $N_1 = 40$,  $N_2 = 50$, $f \sim Beta(2,2.75)$, and $g \sim Beta(2,2)$. Note that $\Delta_{f,g} = 0.10$ with current choice of $f$ and $g$. Let $\theta = 0.1 + 0.01k$, where $k = 0, 1, 2, \cdots, 80$. With a given $\theta$, take $\theta_1 = \theta - 0.1$ and $\theta_2 = \theta + 0.1$.
	\item[(c)] Effect of $\triangle\theta$: $N_1 = 40$, $N_2 = 50$, $f \sim Beta(2,2.75)$, and $g \sim Beta(2,2)$. Note that $\Delta_{f,g} = 0.10$ with current choice of $f$ and $g$. 
	{Let $\triangle \theta = 0.02k$, where $k = 1, 2, \cdots, 45$.}
	With a given $\triangle\theta$, take $\theta_1 = 0.5 - \triangle\theta/2$ and $\theta_2 = 0.5 + \triangle\theta/2$.
	\item[(d)] Effect of sample size: $\theta_1 = 0.3$, $\theta_2 = 0.8$, $f \sim Beta(2,2.75)$, and $g \sim Beta(2,2)$. Note that $\Delta_{f,g} = 0.10$ with current choice of $f$ and $g$. Let $N_2 = 50$, and choose $N_1 \in \{20, 25, 30, \cdots, 115, 120\}$. 
\end{itemize}

We observe that the simulated $\are$ is close to the theoretical $\are$, in terms of both values and the trends as we change the parameters. 
For setting (d), the theoretical line is flat while the simulated line is not flat. The reason is there is an additional term $O(|N_2 - N_1|)$ in the formula, see \eqref{eqn:e1s}. It has a low order compared to $N_1 N_2$, so it was ignored in the main term. The theoretical curve corresponds to the main term, which does not depend on $N_1$, but the simulated curve also takes this remainder term into account. 

\begin{figure}[ht!]
	\centering
	\includegraphics[width=0.98\textwidth]{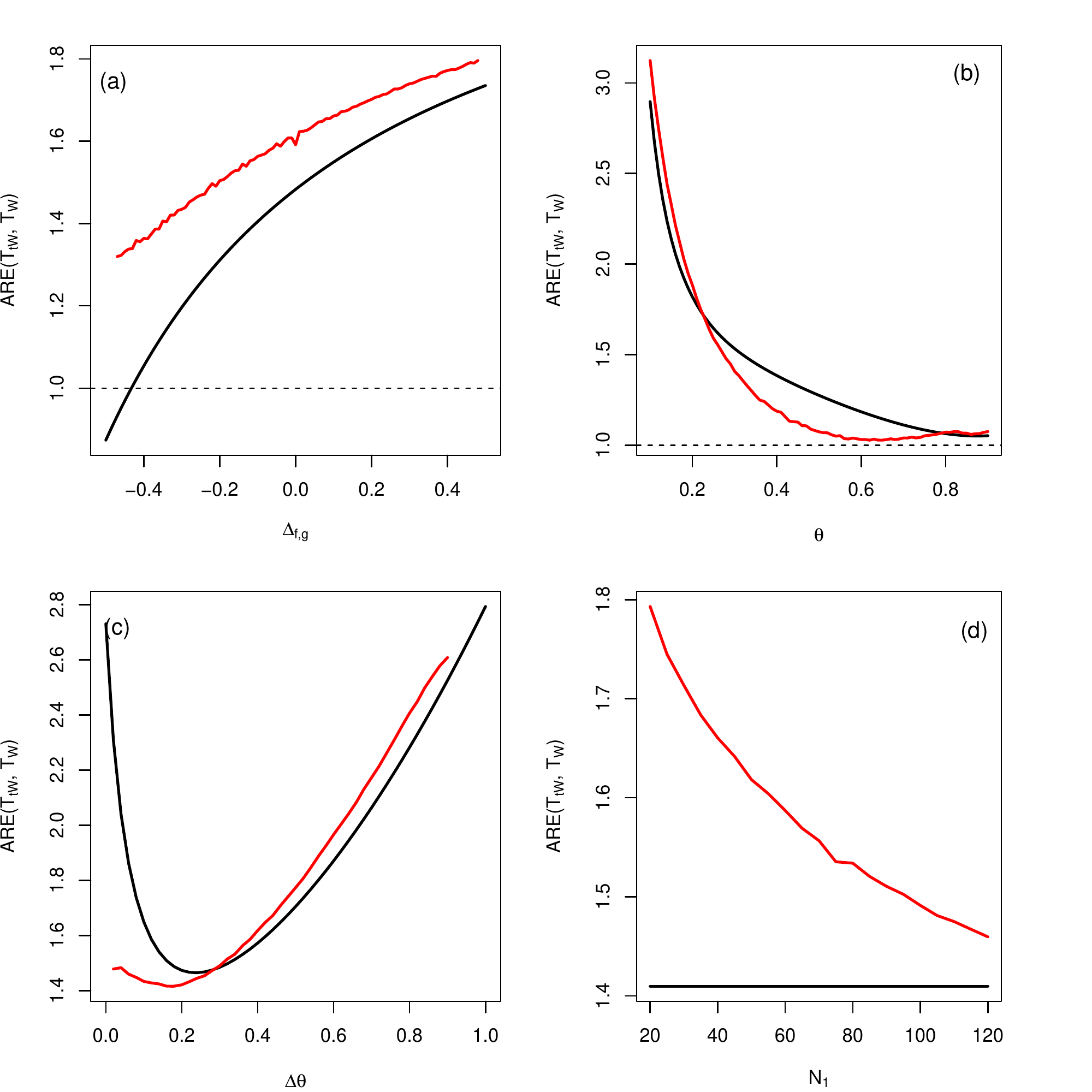}
	\caption{The theoretical (black) and simulated (red) asymptotic  relative efficiency  $\textsc{are}(T_{tW}, T_W)$ comparing the truncated Wilcoxon rank test and the standard Wilcoxon rank test as a function of  (a) 
		$\Delta_{f, g}$, (b)  $\theta$, (c)  $\triangle\theta$,  and (d) $N_1$. {For each plot, the horizontal dashed line represents $\textsc{are}(T_{tW}, T_W)=1$.} }\label{aresimu2}
	
\end{figure}

\section{Basic properties}\label{sec:basic}
Suppose there is a sample with size $n$. Rank all the observations so that the smallest observations have largest ranks. Let $r_i$ and $r_j$ denote the ranks of two randomly selected observations. 
We have the results for the expectation, variance and covariance as follows. The calculations are also presented here although its basic in statistics. 

{\bf Property 1}. $\mathrm{E}[r_i]  = \frac{n+1}{2}$. 

{\bf Property 2}. $\mathrm{E}[r_i^2]  = \frac{(n+1)(2n+1)}{6}$. 

Proof. $\mathrm{E}[r_i^2] = \frac{1}{n} \sum_{i=1}^n i^2 = \frac{(n+1)(2n+1)}{6}$. 

{\bf Property 3}. $\mathrm{Var}[r_i]  = \frac{n^2 - 1}{12}$. 

Proof. Note that $\mathrm{Var}(r_i)  = \mathrm{E}[r_i^2] - (E[r_i])^2$. According to Properties 1--2, the variance is 
$$
\mathrm{Var}(r_i) = \frac{1}{n} \sum_{i=1}^n i^2 - (\frac{n+1}{2})^2 = \frac{n^2 - 1}{12}. 
$$

{\bf Property 4}. $\mathrm{Cov}[r_i, r_j]  = -\frac{n+1}{12}$. 

Proof. When $i \neq j$, we have 
$$
\mathrm{E}[r_i r_j] = \frac{1}{n(n-1)} \sum_{i \neq j} ij = \frac{(n+1)(3n+2)}{12}.
$$
Therefore, 
$$
\mathrm{Cov}(r_i, r_j) = \mathrm{E}[r_i r_j] - [\mathrm{E}(r_i)]^2 = \frac{(n+1)(3n+2)}{12} - (\frac{n+1}{2})^2 = -\frac{n+1}{12}.
$$

{\bf Property 5}. Consider a randomly selected group of observations $S$ with size $n_1$. The rank-sum of this group is denoted by $r_g$. Then, 
$$
\mathrm{E}[r_g] = \frac{1+n}{2} n_1, \qquad \mathrm{Var}(r_g) =  \frac{n_1(n - n_1) (n+1)}{12}. 
$$

Proof. For the expectation part, it simply introduces in $E[r_i]$ for each observation in this group. Next we check the variance. 
\begin{eqnarray*}
	\mathrm{Var}(r_g) & = & \sum_{i\in S} \mathrm{Var}(r_i) + \sum_{i, j \in S, i \neq j} \mathrm{Cov}(r_i, r_j)\\
	& = & \sum_{i\in S} \frac{n^2 - 1}{12} + \sum_{i, j \in S, i \neq j} -\frac{n+1}{12}\\
	& = & \frac{n_1(n-n_1)(n+1)}{12}. 
\end{eqnarray*}

\section{Derivations of  the  Modified  Wilcoxon Rank Sum Statistic}

For a given statistic $S$, we use $\E[S]$  and $\var(S)$ to represent its mean and variance calculated under the null hypothesis and use $\E_1[S]$ to represent its mean calculated under the alternative hypothesis. 
In this section, given that $W = S^2/\var(S)$ where $S = R - \frac{N_1 + N_2 + 1}{2} N_1$, and $s = r - \frac{\lfloor p(N_1 + N_2)\rfloor + 1}{2} \lfloor p N_1 \rfloor - \frac{1}{4} \frac{p_1 + p_2}{2} (1 - \frac{p_1 + p_2}{2})(N_2 - N_1)$, we want to verify the following results:
\begin{itemize}
	\item $\E[S] = 0$, $\E[s] = O(\max\{N_1, N_2\})$;
	\item the results for $\var(S)$ and $\var(s)$: 
	\begin{eqnarray*}
		\mathrm{Var}(S) & = & N_1 N_2 (N_1 + N_2) \theta (1 - \theta + \theta^2/3)/4 + O(N^{2.5}),\\
		\mathrm{Var}(s) &=& N_1 N_2 (N_1 + N_2) \theta^3 (1 - \theta + 1/3)/4 + O(N^{2.5});
	\end{eqnarray*}
\end{itemize}

\vspace{2em}
Proof. Now we prove the results one by one. 
\begin{itemize}
	\item We show the expectation first. The result for $\E[S]$ is direct, so we focus on $\E[s]$. 
	
	Recall that $p = \max\{p_1, p_2\}$, where $p_1 = n_1/N_1$ and $p_2 = n_2/N_2$. 
	Let $\tilde{r} = r  - \frac{\lfloor p (N_1 + N_2) \rfloor + 1}{2} \lfloor pN_1 \rfloor$, and so $s = \tilde{r} - \frac{1}{4} \frac{p_1 + p_2}{2} (1 - \frac{p_1 + p_2}{2}) (N_2 - N_1)$. To find $E[s]$, we need to calculate $E[\tilde{r}]$ and $E[\frac{1}{4} \frac{p_1 + p_2}{2} (1 - \frac{p_1 + p_2}{2}) (N_2 - N_1)]$. 
	
	With independent samples, $p_1 $ and $p_2$  are independent, $\E[p_i] = \theta_i$ and $\var(p_i) = \theta_i(1 - \theta_i)/N_i)$, $i = 1,2$. Under null hypothesis, $\theta_1 = \theta_2 = \theta$. 
	We then have
	\begin{eqnarray}\label{eqn:wil2}
		E\biggl[\frac{1}{4} \frac{p_1 + p_2}{2} (1 - \frac{p_1 + p_2}{2}) (N_2 - N_1)\biggr]
		& = & \frac{N_2 - N_1}{4} \biggl[(\frac{\theta_1 + \theta_2}{2}) - (\frac{\theta_1 + \theta_2}{2})^2 - \frac{\theta_1(1-\theta_1)}{4N_1} - \frac{\theta_2(1-\theta_2)}{4N_2}\biggr] \nonumber\\
		\mbox{(Let } \theta = \frac{\theta_1+\theta_2}{2} ) \qquad & = & \frac{N_2 - N_1}{4} (\theta (1 - \theta)  + O(\frac{1}{N_1} + \frac{1}{N_2})).
	\end{eqnarray}
	Here, $\theta$ is the probability of zero under null and $\theta = (\theta_1 + \theta_2)/2$ under alternative hypothesis. 
	
	Next we calculate $E[\tilde{r}]$. Recall that $r$ is the sum-rank of the truncated sample 1. Let $r_0$ be the sum-rank of all the non-zero elements in the truncated sample 1. With $r_0$, we rewrite $\tilde{r}$ as 
	\begin{eqnarray*}
		\tilde{r} & = & r_0  - \frac{n_1 + n_2 + 1}{2} n_1 + \frac{n_1 + n_2 + 1 + \lfloor p(N_1 + N_2) \rfloor}{2} (\lfloor pN_1 \rfloor - n_1) \\
		&&- \frac{1 + \lfloor p(N_1 + N_2) \rfloor}{2} \lfloor pN_1 \rfloor 
		+ \frac{n_1 + n_2 + 1}{2} n_1 \\
		& = &  (r_0  - \frac{n_1 + n_2 + 1}{2} n_1) + \frac{1}{2} p N_1 N_2 (p_2 - p_1) + O(\max\{N_1, N_2\}),
	\end{eqnarray*}
	where $O(N_2)$ comes from the difference between $\lfloor p N_1\rfloor$ ($\lfloor p N_1 \rfloor$) and $pN_1$ ($p N_2$). 
	
	Since all the observations are non-negative, so the non-zero elements always enjoy the ranks from 1 to $n_1 + n_2$. 
	Hence, under null hypothesis that $f = g$, $E[r_0  - \frac{n_1 + n_2 + 1}{2} n_1] = 0$. Therefore, 
	\begin{equation}\label{eqn:wilr}
		\E[\tilde{r}|p_1, p_2]  = \frac{p N_1 N_2}{2} (p_2 - p_1) + O(\max\{N_1, N_2\}).
	\end{equation}
	We still need to remove the condition that $p_1$ and $p_2$ are given. 
	The following lemma is used to describe the distribution of $p (p_2 - p_1)$.
	\begin{lemx}\label{lemma:willemma}
		Define  $p_1 = n_1/N_1$, $p_2 = n_2/N_2$, and $p = \max\{p_1, p_2\}$. Under null hypothesis that $\theta_1 = \theta_2 = \theta$, $F = G$, and let $N_1, N_2 \rightarrow \infty$ at the same order, there is 
		\begin{eqnarray*}
			&& E[p(p_2 - p_1)/2] = \frac{\theta(1 - \theta)}{4} \frac{N_1 - N_2}{N_1 N_2},\\
			&& E[p^2(p_2 - p_1)^2] = \theta^3 (1 - \theta) (1/N_1 + 1/N_2) + O(N_1^{-3/2}).
		\end{eqnarray*}
	\end{lemx}
	
	Combining Lemma \ref{lemma:willemma} with (\ref{eqn:wilr}) leads to 
	\[
	E[\tilde{r}] =  \frac{\theta(1 - \theta)}{4} (N_1 - N_2) + O(\max\{N_1, N_2\}).
	\]
	Combining this with (\ref{eqn:wil2}), we have that 
	\[
	E[s] = O(\max\{N_1, N_2\}). 
	\]
	The first conclusion is proved. 
	
	\item Next we want to check $\var(S)$ and $\var(s)$. 
	
	Note that $s =  \tilde{r} - \frac{1}{4} \frac{p_1 + p_2}{2} (1 - \frac{p_1 + p_2}{2}) (N_2 - N_1) = \tilde{r} - I$. Therefore, 
	\begin{equation}\label{eqn:vars1}
		\var(s) = \var(\tilde{r}) + \var(I) - 2 \cov(\tilde{r}, I). 
	\end{equation}
	Check the terms one by one. 
	
	The first term is $\var(\tilde{r})$. Note that we already find $\E[\tilde{r}|p_1, p_2] = p N_1 N_2 (p_2 - p_1)/2 + O(\max\{N_1, N_2\})$. Therefore, 
	\begin{eqnarray*}
		\var(\tilde{r}) & = & \var(\E[\tilde{r}|p_1, p_2]) + \E[\var(\tilde{r}|p_1, p_2)]\\
		& = & \frac{N_1^2 N_2^2}{4}\{ E[(p_2 - p_1)^2 p^2] - (E[(p_2 - p_1)p])^2\} + \frac{N_1 N_2}{12} E[N_1 p_1^2 p_2 + N_2 p_1 p_2^2 + p_1 p_2].
	\end{eqnarray*}
	With Lemma \ref{lemma:willemma} we have that 
	\begin{eqnarray*}
		&&E[(p_2 - p_1)^2 p^2] = \theta^3(1 - \theta)(1/N_1 + 1/N_2) + O(N_1^{-3/2}),\\
		&& (E[(p_2 - p_1)p])^2 = \theta^2 (1 - \theta)^2 (N_1 - N_2)^2/ (4 N_1^2 N_2^2 ) = O(N_1^{-2}).
	\end{eqnarray*}
	Combining with $E[N_1 p_1^2 p_2 + N_2 p_1 p_2^2 + p_1 p_2] = (N_1 + N_2) \theta^3 + O(1)$ from basic statistics, we have that 
	\begin{equation}\label{eqn:vartr}
		\var(\tilde{r}) = N_1 N_2 (N_1 + N_2) \theta^3 (1 - \theta + 1/3)/4 + O(\sqrt{N_1} N_2^2).
	\end{equation}
	
	For the second term, note that 
	$$
	\var(I) \leq E[I^2] \leq (N_2 - N_1)^2 E[(\frac{1}{4} \frac{p_1 + p_2}{2} (1 - \frac{p_1 + p_2}{2}))^2] \leq (N_2 - N_1)^2 = O(N^2),
	$$
	and 
	$$
	\cov(\tilde{r}, I) \leq \sqrt{\var(\tilde{r}), \var(I)} \leq (N_2 - N_1) \sqrt{N_1 N_2 (N_1 + N_2) } = O(N^{5/2}). 
	$$
	Introduce the result and (\ref{eqn:vartr}) into (\ref{eqn:vars1}), we have 
	\begin{equation}\label{eqn:vars}
		\var(s) = \var(\tilde{r}) + \var(I) - 2 \cov(\tilde{r}, I) = N_1 N_2 (N_1 + N_2) \theta^3 (1 - \theta + 1/3)/4 + O(N^{2.5}). 
	\end{equation}
	
	Similarly, we have that $\var(S) = N_1 N_2 (N_1 + N_2) \theta (1 - \theta + \theta^2/3)/4 + O(N^{2.5})$.

\end{itemize}

\section{Proof of Theorem \ref{thm:arewilcoxon} in main paper}
Recall that the definition of $\are(W_m, W)$ is 
\[
\are(W_m, W) = \frac{\E[s^2]/\var(s)}{\E_1[S^2]/\var(S)},
\]
where $\E_1$ denote the expectation under alternative hypothesis, and $\var(s)$ means the variance of $s$ under null hypothesis. 
With previous analysis, $\var(s)$ and $\var(S)$ are known. The remaining problem is $\E_1[s^2]$, for which we can derive $\E_1[s]$ and $\var_1(s)$ separately, and the combine them together. Here, $\var_1(s)$ denote the variance of $s$ under alternative hypothesis. 

Recall that we rewrite $s = \tilde{r} - I$, where $\tilde{r}$ can be rewritten as 
\begin{equation*}
	\tilde{r} =  [r_0  - \frac{n_1 + n_2 + 1}{2} n_1]+ \frac{1}{2} p N_1 N_2 (p_2 - p_1) + O(N_2).
\end{equation*}
Since $r_0$ depend on the non-zero part only, so we only consider the non-zero elements of $\mx$ and $\my$. 
Let $\mx(j)$ and $\my(j)$ denote the $j$th non-zero observation in $\mx$ and $\my$, respectively. 
The rank of $\mx(j)$ equals to 
$$
r_{j}= \frac{1}{2} +  \sum_{m = 1}^{n_1} 1\{\mx(j) < \mx(m)\} +  \sum_{m = 1}^{n_2} 1\{\mx(j) < \my(m)\} + \frac{1}{2} \sum_{m = 1}^{n_1} 1\{\mx(j) = \mx(m)\} + \frac{1}{2}\sum_{m = 1}^{n_2} 1\{\mx(j) = \mx(m)\}.
$$
Note that $\E[1\{\mx(j) < \mx(m)\}] = P(X_1 < X_2)$ when $X_1, X_2 \stackrel{i.i.d.}{\sim} f$, and $E[1\{\mx(j) < \my(m)\}] = P(X < Y)$ when $X \sim f, Y \sim g$ independently. 
Therefore, we have
$$\E_1[r_{j}|n_1, n_2] = 1/2 + n_1 P(X_1 < X_2) + n_2 P(X < Y) + \frac{n_1}{2} P(X_1 = X_2) + \frac{n_2}{2}  P(X = Y).$$ 
Note that $P(X_1 = X_2)/2 + P(X_1 < X_2) = 1/2$ since they follow the same distribution and $P(X = Y)/2 + P(X < Y) = \Delta_{f, g} + 1/2$ according to the definition, 
plug them in  and we have that
\begin{equation*}
	\E_1[r_0|n_1, n_2] = \E_1\biggl[\sum_{j = 1}^{n_1} r_{1j}\biggr] = n_1 \biggl[(n_1 + 1)/2 +  n_2 [\Delta_{f,g} + 1/2] \biggr].
\end{equation*}
Introduce it into the definition of $\tilde{r}$, we have 
$$
\E_1[\tilde{r} | n_1, n_2]  = \E_1[r_0|n_1, n_2]  - \frac{n_1+n_2+1}{2} n_1 + \frac{1}{2} p N_1 N_2 (p_2 - p_1)= n_1 n_2 \Delta_{f, g} + \frac{1}{2} p N_1 N_2 (p_2 - p_1).
$$
Further, let $N_1, N_2 \rightarrow \infty$, and note that $n_1 \sim Binomial(N_1, \theta_1)$, $n_2 \sim Binomial(N_2, \theta_2)$, we have that 
$$
\E_1[\tilde{r}]  = \theta_1\theta_2 N_1 N_2 \Delta_{f, g} + \frac{1}{2} \theta_{(m)} N_1 N_2 \Delta \theta,
$$

For the last term $I$, note that $|\E_1[I]| < N_2 - N_1 = O(N)$. Combining all these terms, we have that 
\begin{equation}\label{eqn:e1s}
	\E_1[s]  = \E_1[\tilde{r}]  - \E[I] =  \theta_1\theta_2 N_1 N_2 \Delta_{f, g} + \frac{1}{2} \theta_{(m)} N_1 N_2 \Delta \theta + O(N).
\end{equation}
Similarly, we have the result for $S$ as 
\begin{eqnarray}
	\E_1[S]  & = & \E_1\biggl[\E_1[r_0|n_1, n_2] + \frac{n_1 + n_2 + 1 + N_1 + N_2}{2}(N_1 - n_1) - \frac{N_1 + N_2 + 1}{2} N_1\biggr]\nonumber\\
	& = & \E_1\bigl[n_1n_2 \Delta_{f,g} + \frac{N_1 n_2 - n_1 N_2}{2} \bigr] \nonumber\\
	& = & \theta_1\theta_2 N_1 N_2 \Delta_{f, g} + \frac{1}{2}  N_1 N_2 \Delta \theta.\label{eqn:e1S}
\end{eqnarray}

Now we consider the variance under alternative hypothesis. We deal with $S$ first. 

Note that $S$ is a function of all the observations $\mx$ and $\my$. To find the variance, we try to bound $P(|S - \mu|\geq t)$ for any $t$ with concentration inequality. 
When one observation of $\mx$ changes, the largest change of $S$ caused by it is $N_2$. When one observation of $\my$ changes, the largest change of $S$ is $N_1$. Mathematically, we have 
\begin{eqnarray*}
	|S(x(1), \cdots, x(k), \cdots, x(N_1), \my) - S(x(1), \cdots, x'(k), \cdots, x(N_1), \my)|  & \leq & N_2,\\
	|S(\mx, y(1), \cdots, y(k), \cdots, y(N_2)) - S(\mx, y(1), \cdots, y'(k), \cdots, y(N_2))|  & \leq & N_1.\\
\end{eqnarray*}
According to McDiarmid's Inequality (\cite{McDiarmid}, Page 206), there is 
\[
P(|S - \E[S]| > t) \leq 2 e^{\frac{-2t^2}{N_1 N_2^2 + N_2 N_1^2}}.
\]
Therefore, 
\begin{eqnarray*}
	\var_1(S) & = & \E[|S - \E[S]|^2] = \int_0^{\infty} P(|S - \E(S)|^2 \geq t) dt\\
	& = & \int_0^{\infty} P(|S - \E(S)| \geq \sqrt{t}) dt\\
	& \leq & \int_0^{\infty}2 e^{\frac{-2t}{N_1 N_2^2 + N_2 N_1^2}} dt\\
	& = & N_1 N_2 (N_1 + N_2). 
\end{eqnarray*}
Therefore, 
\begin{equation}\label{eqn:wilvarS}
	\var_1(S) \leq N_1 N_2 (N_1 + N_2) = O(N^3). 
\end{equation}

Similarly, we derive the result for $s$. When one observation of $\mx$ changes, the change of $s$ cannot be larger than the change of $S$, and so 
\begin{equation}\label{eqn:wilvars}
	\var_1(s) \leq N_1 N_2 (N_1 + N_2) = O(N^3). 
\end{equation}

Combine (\ref{eqn:e1s}), (\ref{eqn:e1S}), (\ref{eqn:wilvarS}) and (\ref{eqn:wilvars}), with the results about variance in the second bullet, we have 
\begin{eqnarray*}
	\are(W_m, W) & = & \frac{\E_1[s^2]/\var(s)}{\E_1[S^2]/\var(S)} = \frac{(\E_1[s])^2 + \var_1(s)}{(\E_1[S])^2 + \var_1(S)} \frac{\var(S)}{\var(s)}\\
	& = & \frac{\bigl(\theta_1\theta_2 N_1 N_2 \Delta_{f, g} + \frac{1}{2} \theta_{(m)} N_1 N_2 \Delta \theta\bigr)^2 + O(N^3)}{\bigl(\theta_1\theta_2 N_1 N_2 \Delta_{f, g} + \frac{1}{2}  N_1 N_2 \Delta \theta\bigr)^2 + O(N^3)} \cdot \frac{N_1 N_2 (N_1 + N_2) \theta (1 - \theta + \theta^2/3)/4 + O(N^{2.5})}{N_1 N_2 (N_1 + N_2) \theta^3 (1 - \theta + 1/3)/4 + O(N^{2.5})}\\
	& = & \frac{\bigl(\theta_1\theta_2  \Delta_{f, g} + \frac{1}{2} \theta_{(m)}  \Delta \theta\bigr)^2 }{\bigl(\theta_1\theta_2  \Delta_{f, g} + \frac{1}{2}   \Delta \theta\bigr)^2 } \cdot \frac{ (1 - \theta + \theta^2/3)}{ \theta^2 (1 - \theta + 1/3)}  + O(N^{-1/2}).
\end{eqnarray*}
The result is proved.

\section{Proof of Lemma \ref{lemma:cov-m} in main paper}

Proof. It is easy to verify that $\mathrm{E}[s_i] = 0$, $1 \leq i \leq K$. Since $U_i$ is a linear combination of $s_i$, there is 
$$
E[U_i] = \sum_{j = 1}^{i} E[s_j] - i E[s_{i+1}] = 0, \qquad 1 \leq i \leq K.
$$
To check the variance, we begin with the variances and covariances of $s_i$. Given $n_1, \cdots, n_K$, $\var(s_i|n_1, \cdots, n_K)$ equals to the variance of the sum-rank of non-zero elements, which is already derived in the basic property section in Supplementary materials. Therefore, with the law of total variance, we find that 
\begin{eqnarray*}
	\mathrm{Var}(s_i) &=& \mathrm{E}\biggl[n_i (\sum_{j = 1, j \neq i}^{K} n_j) (\sum_{j = 1}^K n_j + 1)\biggr]/12 + \mathrm{E}\biggl[n^2 (\sum_{j = 1}^K n_j - Kn_i)^2\biggr]/4, \\
	\mathrm{Cov}(s_i, s_j) &=& - \mathrm{E}\biggl[n_i  n_j (\sum_{k = 1}^K n_k + 1)\biggr]/12 + \mathrm{E}\biggl[n^2 (\sum_{k = 1}^K n_k - Kn_i) (\sum_{k = 1}^K n_k - Kn_j)\biggr]/4, 
\end{eqnarray*}
where  $1 \leq i, j \leq K - 1, i \neq j$. From this, we  obtain 
\begin{eqnarray*}
	\mathrm{Var}(U_i) &= & E \biggl\{\frac{\sum_{j = 1}^K n_j + 1}{12} \bigl[(\sum_{j = 1}^{i} n_j) n_{i+1} (i+1)^2 + (\sum_{k = i + 2}^{K} n_k)(\sum_{j = 1}^{i} n_j + i^2 n_{i+1})\bigr] \biggr\}\\
	&& + \mathrm{E}\biggl[\frac{K^2 n^2}{4} (\sum_{j = 1}^{i } n_j - i n_{i+1})^2\biggr] ,\\
	\mathrm{Cov}(U_{i_1}, U_{i_2}) &=& \mathrm{E}\biggl[K^2 n^2 (\sum_{k = 1}^{i_1} n_k - i_1 n_{i_1 + 1}) (\sum_{j = 1}^{i_2} n_j - i_2 n_{i_2 + 1})\biggr]/4,   
\end{eqnarray*}
where   $1 \leq i \leq K - 1$,
$1 \leq i_1 < i_2 \leq K - 1$ and 
$\sum_{k = i + 2}^K n_k = 0$ when $i + 2 > K$. 

Now the remaining term is $ \mathrm{E}\bigl[K^2 n^2 (\sum_{k = 1}^{i_1} n_k - i_1 n_{i_1 + 1}) (\sum_{j = 1}^{i_2} n_j - i_2 n_{i_2 + 1})\bigr]$. We prove the following lemma. 
\begin{lemx}\label{lemma:cov} 
	Let $n_1, \cdots, n_K \stackrel{i.i.d}{\sim} Binomial(N, \theta)$, and $n = \max_{1 \leq i \leq K} n_i$. Given $n$,  we have  
	\begin{equation}\label{eqn:cov}
		\mathrm{E}\biggl[ (\sum_{k = 1}^{i_1} n_k - i_1 n_{i_1 + 1}) (\sum_{j = 1}^{i_2 } n_j - i_2 n_{i_2 + 1})|n\biggr] = 0, \quad 1 \leq i_1 < i_2 \leq K - 1.
	\end{equation}
\end{lemx}
With the lemma, we have that 
$$
\mathrm{Cov}(U_{i_1}, U_{i_2}) = \mathrm{E}\biggl[ K^2 n^2  \E[(\sum_{k = 1}^{i_1} n_k - i_1 n_{i_1 + 1}) (\sum_{j = 1}^{i_2} n_j - i_2 n_{i_2 + 1})|n]\biggr]/4 = \E[K^2 n^2 \cdot 0] = 0.
$$
The result is proved. \qed

Proof of Lemma \ref{lemma:cov}:

To show (\ref{eqn:cov}), note that 
\begin{eqnarray*}
	\E\biggl[ (\sum_{k = 1}^{i_1} n_k - i_1 n_{i_1 + 1}) (\sum_{j = 1}^{i_2 } n_j - i_2 n_{i_2 + 1})|n\biggr] & = & 
	\E\biggl[ \sum_{k = 1}^{i_1} (n_k - n_{i_1 + 1}) \sum_{j = 1}^{i_2 } (n_j -  n_{i_2 + 1})|n\biggr]\\
	& = & \sum_{k = 1}^{i_1}\sum_{j = 1}^{i_2 } \E\biggl[  (n_k - n_{i_1 + 1})  (n_j -  n_{i_2 + 1})|n\biggr],
\end{eqnarray*}
for $1 \leq i_1 < i_2 \leq K - 1$. 
Therefore, to show the left-hand side is 0, we check the value of each single term 
\[
\E[(n_k - n_{i_1}) (n_j - n_{i_2})|n].
\]
As $k \leq i_1 < i_2$, and $1 \leq j \leq i_2 - 1$, there are 3 cases: 1. $j = k$; 2. $j = i_1$; 3. $j \neq k$ and $j \neq i_1$. 

In case 1, we have that 
\[
\E[(n_k - n_{i_1}) (n_k - n_{i_2})|n] = \E[n_k^2 - n_{i_1} n_k - n_{i_2} n_k + n_{i_1} n_{i_2}|n]
\]
Given $n$, $n_i n_j$ has the same joint distribution for any $i \neq j$, and so they share the same expectation. What's more, $n_i$ and $n_j$ has the same conditional variance given $n$. Therefore, 
\begin{eqnarray}\label{eqn:11}
	&&\E[n_k^2 - n_{i_1} n_k - n_{i_2} n_k + n_{i_1} n_{i_2}|n] \nonumber\\
	& = & (\E[n_k|n])^2  + \var(n_k|n) - \E[n_{i_1}|n] \E[n_{k}|n] - \cov(n_{i_1}|n, n_{k}|n)\nonumber\\
	& = &  \var(n_1|n)  - \cov(n_1|n, n_2|n).
\end{eqnarray}
The last equation stands since $n_k|n$ and $n_{i_1}|n$ follow the same distribution. 

In case 2, we have that 
\[
\E[(n_k - n_{i_1}) (n_{i_1} - n_{i_2})|n] = \E[n_k n_{i_1} - n_{i_1}^2 - n_{i_2} n_k + n_{i_1} n_{i_2}|n].
\]
With the same analysis of (\ref{eqn:11}), we have that 
\begin{equation}\label{eqn:12}
	\E[n_k n_{i_1} - n_{i_1}^2 - n_{i_2} n_k + n_{i_1} n_{i_2}|n] =\cov(n_1|n, n_2|n) -  \var(n_k|n).
\end{equation}

In case 3, there is 
\begin{equation}\label{eqn:13}
	\E[(n_k - n_{i_1}) (n_j - n_{i_2})|n] = \E[n_k n_j - n_{i_1} n_j - n_{i_2} n_k + n_{i_1} n_{i_2}|n] = 0.
\end{equation}

Combining (\ref{eqn:11})--(\ref{eqn:13}), and summation over $1 \leq k \leq i_1 $ and $1 \leq j \leq i_2 $, we can find that the mean is 0, and (\ref{eqn:cov}) follows.
\qed

\section{Proof of Lemma \ref{lemma:varu-m} in main paper}
In this proof,  we use the  calculation of $\var(U_{K - 1})$ as an example for simplicity. The calculation of $\var(U_i)$ is similar. 

We first consider the term under null hypothesis. 
According to (\ref{eqn:varsame}), there is an expression for $\var(U_{K-1})$. Decompose $\var(U_{K-1})$ into two parts,
\begin{equation*}\label{varpf}
	\frac{\var(U_{K-1})}{K^2} =  \frac{\E[n_K (\sum_{j = 1}^{K-1} n_j) (\sum_{j = 1}^K n_j + 1)]}{12} + \frac{\E[n^2 (\sum_{j = 1}^K n_j - Kn_K)^2]}{4} = I + II.
\end{equation*}
Part $I$ can be easily calculated by the distribution of $n_j$, and with basic calculations we obtain 
\begin{equation}\label{varpf1}
	I = N^3\theta^3 K (K - 1)/12 + O(N^2).
\end{equation}
For $II$, we take $n = m + N\theta$, where $m = \max\{n_1 - N\theta, n_2 - N\theta, \cdots, n_K - N \theta\}$. Then we have 
\begin{eqnarray*}
	4 \cdot II &=& \E[(m + N\theta)^2  (\sum_{j = 1}^K n_j - Kn_K)^2]\\
	&  =& N^2 \theta^2 \E[(\sum_{j = 1}^K n_j - Kn_K)^2] +  2N \theta \E[m (\sum_{j = 1}^K n_j - Kn_K)^2] +  \E[m^2(\sum_{j = 1}^K n_j - Kn_K)^2]\\
	& = & IIa + IIb + IIc.
\end{eqnarray*}
With basic calculations and that $(\sum_{j = 1}^K n_j - Kn_K)/N$ is asymptotically distributed as a normal distribution with mean 0 and variance $K(K - 1) \theta (1 - \theta) $, we have that 
\begin{equation}\label{eqn:varpf2a}
	IIa = N^3 \theta^3 K(K-1) (1 - \theta) + O(N^2).
\end{equation}
For $IIb$ and $IIc$, according to Cauchy-Schwarz inequality, we have that 
\begin{equation}\label{eqn:varpfIIb}
	IIb \leq 2N\theta \sqrt{\E[m^2] \E[(\sum_{j = 1}^K n_j - Kn_K)^4]} = 2N \theta \sqrt{\E[m^2]} \sqrt{3} K (K - 1) N\theta(1 - \theta),
\end{equation}
and
\begin{equation}\label{eqn:varpfIIc}
	IIc \leq \sqrt{\E[m^4] \E[(\sum_{j = 1}^K n_j - Kn_K)^4]} = \sqrt{\E[m^4]} \sqrt{3} K (K - 1) N\theta(1 - \theta).
\end{equation}
The unknown part here is $\E[m^2]$ and $\E[m^4]$. For these two terms, we can bound them by  
\begin{eqnarray*}
	&\E[m^4] &\leq K \E[(n_1 - N\theta)^4] = O(N^2), \\
	&\E[m^2] &\leq K \E[(n_1 - N\theta)^2] = O(N).
\end{eqnarray*}
Plugging these  into (\ref{eqn:varpfIIb}) and (\ref{eqn:varpfIIc}), we have that
\begin{equation}\label{eqn:varpf2bc}
	IIb \lesssim O(N^{5/2}), \qquad IIc \lesssim O(N^2).
\end{equation}
Combining (\ref{eqn:varpf2a}) and (\ref{eqn:varpf2bc}), we have that
\[
II = N^3 \theta^3 K(K - 1) (1 - \theta)/4 + O(N^{5/2}).
\]
Combining this with (\ref{varpf1}), and we have that
\[
\var(U_{K-1}) = N^3 \theta^3 K^3 (K - 1)(1/3 + (1 - \theta))/4 + O(N^{5/2}).
\]
The result under null hypothesis is proved. 

Next, we consider the variances under alternative hypothesis. The analysis is similar with what we did for the modified Wilcoxon statistics under alternative. 

Consider $Y_i$ first. $Y_i$ is a function of $\mx_1, \mx_2, \cdots, \mx_K$. When one observation change, the change of $Y_i$ is no larger than $K N$. For $U_i$, the same thing happens that the change is no larger than $KN$. 
Therefore, 
\[
\sum_{i=1}^{KN} (KN)^2  = K^3 N^3. 
\]
According to McDiarmid's Inequality (\cite{McDiarmid}, Page 206), there is 
\[
P(|Y_i - \E[Y_i]| > t) \leq 2 e^{\frac{-2t^2}{K^3 N^3}}.
\]
Therefore, 
\begin{eqnarray*}
	\var_1(Y_i) & = & \E[|Y_i - \E[Y_i]|^2] = \int_0^{\infty} P(|Y_i - \E(Y_i)|^2 \geq t) dt\\
	& = & \int_0^{\infty} P(|Y_i - \E(Y_i)| \geq \sqrt{t}) dt\\
	& \leq & \int_0^{\infty}2 e^{\frac{-2t}{K^3 N^3}} dt\\
	& = & K^3 N^3. 
\end{eqnarray*}
Therefore, 
\begin{equation}\label{eqn:modkwvar1s}
	\var_1(Y_i) \leq K^3 N^3, \quad  \var_1(U_i) \leq K^3 N^3, \qquad 1 \leq i \leq K.
\end{equation}

The lemma is proved.\qed

\section{Proof of Lemma \ref{lemma:eu} in main paper}
\setcounter{equation}{0}

In this section, all the proofs are for the expectations under alternative hypothesis. So we drop the subscriptsof $\E_1$ for simplicity in this section only.

With basic calculations, it can be shown that, for $1 \leq i \leq K - 1$,
\[
Y_i = (\sum_{j = 1}^i s_j^0 - i s_{i+1}^0) + \frac{NK}{2}[ i n_{i+1} - \sum_{j = 1}^i n_j], 
\;
U_i = (\sum_{j = 1}^i s_j^0 - i s_{i+1}^0) + \frac{n K}{2}[ i n_{i+1} - \sum_{j = 1}^i n_j].
\]
So, for $Y_i$ terms, the problem reduces to find $\E[\frac{NK}{2}( i n_{i+1} - \sum_{j = 1}^i n_j)]$. With basic calculations, this expectation turns to be $\frac{N^2 K}{2}( i \theta_{i+1} - \sum_{j = 1}^i \theta_j)$. Combining with $\E\bigl[\sum_{j = 1}^{i} s_j^0 - i s_{i + 1}^0 \big| n_1, \cdots, n_K\bigr] =  \sum_{j = 1}^i \sum_{k \neq j} n_j n_k \Delta_{j, k} - i n_{i+1}  \sum_{k \neq i+1} n_k \Delta_{i+1, k}$, we have that
\[
\E[Y_i] = N^2 \biggl[ \sum_{j = 1}^i \sum_{k \neq j} \theta_j \theta_k \Delta_{j, k} - i \theta_{i+1}  \sum_{k \neq i+1} \theta_k \Delta_{i+1, k} \biggr] + \frac{KN^2}{2} \biggl(i \theta_{i+1} - \sum_{j = 1}^i \theta_{j}\biggr), \quad 1 \leq i \leq K - 1.
\]
So, the first part of the conclusion is proved. 

To find $\E[U_i]$, it also reduces to find $\E_1[\frac{nK}{2}[ i n_{i+1} - \sum_{j = 1}^i n_j]]$. Decompose $n = N \theta_{(K)} + m$, then we have that 
\begin{eqnarray*}
	\E[U_i] &=& N^2 \biggl[ \sum_{j = 1}^i \sum_{k \neq j} \theta_j \theta_k \Delta_{j, k} - i \theta_{i+1}  \sum_{k \neq i+1} \theta_k \Delta_{i+1, k} \biggr]\\
	&& + \frac{KN^2}{2} \theta_{(K)}(i \theta_{i+1} - \sum_{j = 1}^i \theta_{j}) + \E[\frac{mK}{2}[ i n_{i+1} - \sum_{j = 1}^i n_j]].
\end{eqnarray*}
As long as we can show that $\E[\frac{mK}{2}[ i n_{i+1} - \sum_{j = 1}^i n_j]] = o(N^2)$, the lemma is proved. 

With Cauchy-Schwarz Inequality, we have that 
\begin{equation}\label{eqn:exppf1}
	\E[\frac{mK}{2}[ i n_{i+1} - \sum_{j = 1}^i n_j]] \leq \frac{K}{2}\sqrt{\E[m^2] \E[( i n_{i+1} - \sum_{j = 1}^i n_j)^2]}.
\end{equation}

If $\theta_1 = \theta_2 = \cdots = \theta_K = \theta$, then $\theta_{(K)} = \theta$. What's more, $\E[m^2] \leq \sum_{i = 1}^K \E[(n_i - N \theta)^2] = K N \theta (1 - \theta) = O(N)$. The second term that $\E[( i n_{i+1} - \sum_{j = 1}^i n_j)^2] = i(i+1)N\theta (1 - \theta) = O(N)$. 
Combining the two terms with (\ref{eqn:exppf1}), it can be concluded that $\E[\frac{mK}{2}[ i n_{i+1} - \sum_{j = 1}^i n_j]] \leq O(N)$. So, the result is proved.

On the other hand, if there is some $i$ and $j$, such that $\theta_i \neq \theta_j$. Then, 
\begin{equation}\label{eqn:exppf2}
	\E[( i n_{i+1} - \sum_{j = 1}^i n_j)^2] \lesssim O(N^2).
\end{equation} 
For any given $N$, we decompose the index set $S = \{1, 2, \cdots, K\}$ into $S_1 = \{j: \theta_j - \theta_{(K)} \geq - 2\sqrt{\log(N)/N}\}$ and $S_2 = S\backslash S_1$. 
Correspondingly, assume that the index that achieves the maximum is $h$, then the expectation can also be decomposed as following
\begin{equation*}\label{eqn:exppf3}
	\E[m^2] = \E[m^2 1\{h \in S_1\}] + \E[m^2 1\{h \in S_2\}] = I + II.
\end{equation*}
For part $I$, we have that 
\begin{equation}\label{eqn:exppfI}
	I \leq \E[\sum_{j \in S_1} n_j^2] \leq  4\log(N) N + N/4 = O(N \log(N)).
\end{equation}
To bound $II$, assume that $\theta_1 = \theta_{(K)}$ without loss of generality. Then we have that 
\begin{equation}\label{eqn:exppfIIa}
	II \leq \sum_{j \in S_2} \E[(n_j - N \theta_{(K)})^2 1\{n_j  \geq n_1\}] \leq \sum_{j \in S_2} \sqrt{\E[(n_j - N \theta_{(K)})^4] P(n_j \geq n_1)}.
\end{equation}
The last inequality comes from Cauchy-Schwartz Inequality. 
Recall that we have $\theta_j - \theta_{(K)} < - 2\sqrt{\log(N)/N}\}$ as $j \in S_2$. According to Hoeffding's Inequality, we have that 
\[
P(n_j - n_1 \geq 0) \leq P((n_j - n_1) - N(\theta_j - \theta_{(K)}) >  2\sqrt{\log(N) N}\}) \leq \exp(-\frac{2 (2\sqrt{\log(N) N})^2 }{4N}),
\]
and it reduces to $P(n_j - n_1 \geq 0) \leq 1/N^2$.
Combining with (\ref{eqn:exppfIIa}), we have that 
\[
II \leq O(\sqrt{1/N^2 * N^4}) = O(N).
\]
Combining this result about $II$ with (\ref{eqn:exppfI}), there is $\E[m^2] = O(N \log(N))$. Plugging  this result and (\ref{eqn:exppf2}) into (\ref{eqn:exppf1}),  we have 
\[
\E[\frac{mK}{2}[ i n_{i+1} - \sum_{j = 1}^i n_j]] = O(N^{3/2} \sqrt{\log(N)}) = o(N^2).
\]
So, the lemma is proved.
\qed

\par

\section{Proofs of Lemma \ref{lemma:willemma} in Supplemental Materials}
\setcounter{equation}{0}

When $N_1, N_2 \rightarrow \infty$, according to Central Limit Theorem, we have that 
\[
\left(\begin{array}{c}
	p_1\\
	p_2
\end{array}
\right)
\stackrel{(d)}{\rightarrow}
N \left(
\left(\begin{array}{c}
	\theta\\
	\theta
\end{array}
\right)
\left(\begin{array}{cc}
	\theta(1 - \theta)/N_1 & 0\\
	0 & \theta(1 - \theta)/N_2
\end{array}
\right)
\right).
\]
Let $\alpha = p_1 - p_2$, $\beta = (N_1 p_1 + N_2 p_2)/(N_1 + N_2)$, inversely there are $p_1 = \beta + \frac{N_2}{N_1 + N_2} \alpha$ and $p_2 = \beta - \frac{N_1}{N_1 + N_2} \alpha$. 
What's more, the distribution of $(\alpha, \beta)^T$ is that
\[
\left(\begin{array}{c}
	\alpha\\
	\beta
\end{array}
\right)
=
N \left(
\left(\begin{array}{c}
	0\\
	\theta
\end{array}
\right)
\left(\begin{array}{cc}
	\theta(1 - \theta)(1/N_1 + 1/N_2) & 0\\
	0 & \theta(1 - \theta)/(N_1 + N_2)
\end{array}
\right)
\right).
\]
With $\alpha$, $\beta$, we rewrite the term $p(p_2 - p_1)$ as 
\[
p(p_2 - p_1) = 
\left\{
\begin{array}{cc}
	\frac{1}{2}(\beta + \frac{N_2}{N_1 + N_2} \alpha)(-\alpha) & \qquad \alpha \geq 0\\
	\frac{1}{2}(\beta - \frac{N_1}{N_1 + N_2} \alpha)(-\alpha) & \qquad \alpha < 0\\
\end{array}
\right.
\]
As $\alpha \perp \beta$ asymptotically, we have 
\[
E[\alpha \beta] = E[\alpha] E[\beta] = 0, \quad 
E[\alpha^2 1\{\alpha \geq 0\}] = E[\alpha^2 1\{\alpha < 0\}] = \frac{1}{2}\theta(1 - \theta)(1/N_1 + 1/N_2).
\]
Combining with these and basic calculations, we have that 
\[
E[p(p_2 - p_1)] = \frac{\theta(1 - \theta)}{4}\frac{N_1 - N_2}{N_1 N_2}.
\]

On the other hand, assume $N_1 \leq N_2$ without loss of generality, we have that 
\begin{eqnarray*}
	&&E[\alpha^2 \beta^2] = \theta^3(1 - \theta)(1/N_1 + 1/N_2) + O(N_2^{-2}),\\
	&&E[\alpha^3 1\{\alpha \geq 0\}] = E[\alpha^3 1\{\alpha < 0\}] = O(N_1^{-3/2}),\\
	&&E[\alpha^4 1\{\alpha \geq 0\}] = E[\alpha^4 1\{\alpha < 0\}] = O(N_1^{-2}).
\end{eqnarray*}
So, we have that $E[p^2(p_2 - p_1)^2] = E[\alpha^2 \beta^2] + O(E[\alpha^3 \beta]) + O(E[\alpha^4]) = \theta^3 (1 - \theta) (1/N_1 + 1/N_2) + O(N_1^{-3/2})$

\section{Relevant Lemmas for truncated Kruskal-Wallis test for unequal sample sizes}
We analyze $\E[U_i]$ under null hypothesis first. It is a linear function of $s_i$'s. 
For $s_i$, given the number of non-zero entries in each sample (equivalently, $p_1, \cdots, p_K$ are given), there is 
\begin{equation*}\label{eqn:means}
	\mathrm{E}[s_i|p_1, p_2, \cdots, p_K] = \frac{N_i}{2} \biggl[p \sum_{j = 1}^{K} (p_j - p_i) N_j\biggr] + O\biggl(\max\{N_i, \sum_{i = 1}^{K} n_i\}\biggr), 
\end{equation*}
where the remainder comes from the fact that we take the largest integer smaller than $p N_i$ instead of $p N_i$ for each sample. 
Therefore, 
\begin{equation}\label{eqn:eucond}
	\mathrm{E}[U_i] =\frac{1}{2} N_{i+1} \sum_{k = 1}^{K} N_k \sum_{j = 1}^{i } N_j \mathrm{E}[p(p_{i+1} - p_j)] + 
	O\biggl(\sum_{k = 1}^{K} N_k \sum_{j = 1}^{i + 1} N_j\biggr), \qquad 1 \leq i \leq K - 1.
\end{equation}
The expectation in above equation is hard to calculate. However, we  cn show it is comparatively small, hence we only need to find an upper bound, which is given  as Lemma \ref{lemma:expbound} (see Supplementary Materials).
\begin{lemx}\label{lemma:expbound}
	Let $p_i \sim N(\theta, \theta(1 - \theta)/N_i)$ independently, $i = 1, \cdots, K$, and $N_{(1)} = \min\limits_{1 \leq i \leq K} N_i$,  then 
	\[
	| \mathrm{E}[p(p_j - p_i)] | \leq (2\sqrt{K} + 1)\theta(1 - \theta)/N_{(1)}, \qquad i \neq j, 1 \leq i,j \leq K,
	\]
	where $p = \max\limits_{1 \leq i \leq K} p_i$.
\end{lemx}
Based on Lemma \ref{lemma:expbound},  let $N_{(1)} = \min\limits_{1 \leq i \leq K} N_i$, 
\begin{equation}\label{eqn:kwdiffeu}
	\mathrm{E}[U_i] = O\biggl(N_{i+1} \sum\limits_{k = 1}^{K} N_k \sum\limits_{j = 1}^{i } N_j/N_{(1)}\biggr),  1 \leq i \leq K - 1.
\end{equation}

Next, we derive the order of $\mathrm{Var}(U_i)$. It is given by 
\begin{eqnarray}\label{eqn:anovavar}
	\mathrm{Var}(U_i) & = & \mathrm{Var}(\mathrm{E}[U_i|p_1, \cdots, p_K]) + \mathrm{E}[\mathrm{Var}(U_i|p_1, \cdots, p_K)]\\ \nonumber
	& \geq & \mathrm{E}[\mathrm{Var}(U_i|p_1, \cdots, p_K)], \qquad 1 \leq i \leq K - 1\nonumber
\end{eqnarray}
\begin{lemx}\label{lemma:diffvaru}
	Under the null hypothesis that $f_1 = f_2 = \cdots = f_K = f$, there is
	\begin{gather*}\label{eqn:anovavar1}
		\mathrm{Var}[U_i|p_1, \cdots, p_K] =\\
		\frac{1}{12}\mathrm{E}\biggl\{ (\sum\limits_{k = 1}^{K} n_k + 1) \bigl[\sum\limits_{k = 1}^{K} n_k [N_{i+1}^2 \sum\limits_{j = 1}^{i} n_j + n_{i+1} (\sum\limits_{j = 1}^{i} N_j)^2] - (n_{i+1}\sum\limits_{j = 1}^{i} N_j - N_{i+1}\sum\limits_{j = 1}^{i} n_j)^2 \bigr] \biggr\}. \nonumber
	\end{gather*}
	Further, the expectation follows that 
	\begin{eqnarray*}
		\mathrm{E}(\mathrm{Var}[U_i|p_1, \cdots, p_K] ) & = & \frac{\theta^2}{12} N_{i+1} (\sum_{j=1}^i N_j)(\sum_{j=1}^{i+1} N_j)(\sum_{j=1}^K N_j) [(\sum_{j=1}^K N_j)\theta + 3 - 2\theta]\\
		& \geq &  \frac{\theta^3}{12} N_{i+1} (\sum_{j=1}^i N_j)^2(\sum_{j=1}^K N_j)^2.
	\end{eqnarray*}
\end{lemx}

Combine (\ref{eqn:anovavar}) with Lemma \ref{lemma:diffvaru}, the lower bound for $\mathrm{Var}(U_i)$ is 
\begin{eqnarray}\label{lower.bound}
	\mathrm{Var}(U_i) \geq \frac{\theta^3}{12} N_{i+1} (\sum_{j=1}^i N_j)^2(\sum_{j=1}^K N_j)^2  = O\biggl( N_{i+1}^2 (\sum_{j = 1}^{K} N_j)^2 \biggl(\sum_{j = 1}^{i} N_j\biggr)^2 \theta^3 /N_{i+1} \biggr), 
\end{eqnarray}
when $1 \leq i \leq K$.
Compare it to the $E[U_i]$, where $(E[U_i])^2 = O\biggl(N_{i+1}^2 (\sum_{j = 1}^{K} N_j)^2 (\sum_{j = 1}^{i} N_j)^2/N_{(1)}^2\biggr)$. $E[U_i]/\mathrm{Var}(U_i) = o(1)$ when $\max_{1 \leq i \leq K} N_i/N_{(1)}^2 \rightarrow 0$ and $N_{(1)} \rightarrow \infty$.  

For the covariance term, we apply the following lemma. 
\begin{lemx}\label{lemma:diffcov}
	Under the null hypothesis where $f_1 = f_2 = \cdots = f_K = f$, and all $\theta_i$'s are equal, there is 
	\begin{eqnarray*}
		|\mathrm{Cov}(U_i, U_j)|  \leq \theta^2 (\sum_{k = 1}^K N_k)^2 N_{i+1} N_{j+1}(1 + \sum_{k=1}^K N_k/N_{(1)}) \sqrt{(\sum_{l = 1}^i N_l)(\sum_{l = j}^i N_l)} \sqrt{(\sum_{k = 1}^K \frac{1}{N_k})},
	\end{eqnarray*}
	where $N_{(1)} = \min_{1 \leq k \leq K} N_k$ and $N_{(1)} \rightarrow \infty$. 
\end{lemx}

\section{Proof of Lemma \ref{lemma:expbound}}
\setcounter{equation}{0}

With Cauchy-Schwarz inequality and Jensen's inequality, note that $f(x) = \sqrt{x}$ is a concave function, we have that
\begin{eqnarray}\label{eqn:bound}
	\bigl|E[p(p_j - p_i)]\bigr| & = & \bigl|E[(p - p_i)(p_j - p_i)] + E[p_i(p_j - p_i)] \bigr| \nonumber\\
	&\leq& \sqrt{E[(p - p_i)^2] E[(p_j - p_i)^2]} + \bigl|E[p_i(p_j - p_i)]\bigr| .
\end{eqnarray}
For the first part, note that $p_j N_j \sim Binomial(N_j, \theta)$, $p_i N_i \sim Binomial(N_i, \theta)$, $N_i$ and $N_j$ are constants and $p_i$ and $p_i$ are independent, so we have 
\begin{equation}\label{eqn:bound1}
	E[(p_j - p_i)^2] = \theta(1 - \theta) (1/N_j + 1/N_i).
\end{equation}
To bound $E[(p - p_i)^2]$, note that we have $(p - p_i)^2 \leq \sum_{k = 1}^K (p_k - p_i)^2$. Combining with (\ref{eqn:bound1}), there is
\begin{equation}\label{eqn:bound2}
	E[(p - p_i)^2] \leq   \sum_{k = 1}^K E[(p_k - p_i)^2] 
	= \sum\limits_{k \neq i} \theta(1 - \theta) (1/N_k + 1/N_i)
	\leq 2K \theta(1 - \theta)/N_{(1)}.
\end{equation}
Introduce (\ref{eqn:bound1}) and (\ref{eqn:bound2}) into (\ref{eqn:bound}), and we have
\[
|E[p(p_j - p_i)]|  \leq 2\theta(1 - \theta) \sqrt{K}/N_{(1)} + \theta(1 - \theta)/N_i .
\]
So the result follows. \qed

\section{Proof of Lemma \ref{lemma:diffvaru}}
\setcounter{equation}{0}

There are two results to prove in this lemma, the conditional variance, and the expectation of the conditional variance. 
\begin{itemize}
	\item Conditional on the number of nonzeros in each sample, the variance is 
	\begin{gather*}
		\mathrm{Var}[U_i|p_1, \cdots, p_K] =\\
		\frac{1}{12}(\sum\limits_{k = 1}^{K} n_k + 1) \bigl[\sum\limits_{k = 1}^{K} n_k [N_{i+1}^2 \sum\limits_{j = 1}^{i} n_j + n_{i+1} (\sum\limits_{j = 1}^{i} N_j)^2] - (n_{i+1}\sum\limits_{j = 1}^{i} N_j - N_{i+1}\sum\limits_{j = 1}^{i} n_j)^2 \bigr]. \nonumber
	\end{gather*}
	\item The expectation of the variance is 
	\begin{eqnarray*}
		\mathrm{E}(\mathrm{Var}[U_i|p_1, \cdots, p_K] ) & = & \frac{\theta^2}{12} N_{i+1} (\sum_{j=1}^i N_j)(\sum_{j=1}^{i+1} N_j)(\sum_{j=1}^K N_j) [(\sum_{j=1}^K N_j)\theta + 3 - 2\theta]\\
		& \geq &  \frac{\theta^3}{12} N_{i+1} (\sum_{j=1}^i N_j)^2(\sum_{j=1}^K N_j)^2.
	\end{eqnarray*}
\end{itemize}
We will show them one by one. 

First we find the conditional variance. Recall that $U_i = \sum_{j=1}^i (N_{i+1}s_j - N_j s_{i+1})$. Define $r_j^0$ as the sum-rank of non-zero elements in Sample $j$. Hence, $s_j = r_j^0 + \frac{ \sum_{i=1}^K \lfloor p N_i \rfloor+ 1 + \sum_{i = 1}^K n_i}{2} (\lfloor p N_i\rfloor - n_i) - \frac{ \sum_{i=1}^K \lfloor p N_i \rfloor+ 1}{2} (\lfloor p N_i\rfloor)$.
Given $p_1, p_2, \cdots, p_K$, equivalently $n_1, n_2, \cdots, n_K$, the only random part is $r_j^0$. Therefore, 
\[
\var(U_i|p_1, \cdots, p_K) = \var(\sum_{j=1}^i (N_{i+1}s_j - N_j s_{i+1})|p_1, \cdots, p_K) =  \var(\sum_{j=1}^i (N_{i+1}r^0_j - N_j r^0_{i+1})|p_1, \cdots, p_K). 
\]  
Given $p_1, \cdots, p_K$, then $n_1, n_2, \cdots, n_K$ are also given. It means we have $\sum_{i=1}^K n_i$ independently and identically distributed observations, and we are interested in the sum-rank of one sample $j$, denoted as $r_j^0$. Definitely, we can use the properties in Section \ref{sec:basic}. 
\begin{eqnarray*}
	&&\var(\sum_{j=1}^i (N_{i+1}r^0_j - N_j r^0_{i+1})|p_1, \cdots, p_K) \\
	& = & \var(N_{i+1} \sum_{j=1}^i r_j^0 - r_{i+1}^0 \sum_{j=1}^i N_j)\\
	& = & N_{i+1}^2 \sum_{j=1}^i \var(r_j^0) + ( \sum_{j=1}^i N_j)^2 \var(r_{i+1}^0) - 2 N_{i+1}  (\sum_{j=1}^i N_j) \sum_{j=1}^i \cov(r_{i+1}^0, r_{j}^0).
\end{eqnarray*}
According to the properties in Section \ref{sec:basic}, $\var(r_i^0) = \frac{(\sum_{k=1}^K n_k + 1)}{12} n_i (\sum_{k\neq i, k = 1}^K n_k)$, $\cov(r_i^0, r_j^0) = -\frac{(\sum_{k=1}^K n_k + 1)}{12} n_i n_j$. Introduce the terms into the equation, and we have that 
\begin{eqnarray*}
	&&\mathrm{Var}[U_i|p_1, \cdots, p_K] \nonumber\\
	& = & \frac{\sum_{k = 1}^{K} n_k + 1}{12} \bigl[N_{i+1}^2 (\sum\limits_{j=1}^i n_j)(\sum\limits_{j=i+1}^K n_j) + (\sum_{j=1}^i N_j)^2 n_{i+1} (\sum_{j \neq i+1,j=1}^K n_j) + 2 N_{i+1} (\sum_{j=1}^i N_j) n_{i+1} (\sum_{j=1}^i n_j)
	\bigr]\\
	& = &
	\frac{1}{12} (\sum\limits_{k = 1}^{K} n_k + 1) \bigl[\sum\limits_{k = 1}^{K} n_k [N_{i+1}^2 \sum\limits_{j = 1}^{i} n_j + n_{i+1} (\sum\limits_{j = 1}^{i} N_j)^2] - (n_{i+1}\sum\limits_{j = 1}^{i} N_j - N_{i+1}\sum\limits_{j = 1}^{i} n_j)^2 \bigr]. \nonumber
\end{eqnarray*}
The first conclusion is proved. 

To derive the second conclusion, we simply calculate the expectation of the results. Note that in the equation, there is no $p$ term, only products of $n_i$, $n_j$ and $n_k$. Under null hypothesis, $n_i \sim Binomial(N_i, \theta)$ independently, therefore, we have 
\begin{eqnarray*}
	&&\E[n_i] = N_i \theta, \quad \var(n_i) = N_i \theta(1 - \theta), \quad \E[n_i^2] = N_i\theta(N_i \theta + 1-\theta), \\
	&&\E[n_i n_j] = N_i N_j \theta^2, \quad \E[n_i^2 n_j] = N_iN_j\theta^2(N_i\theta + 1-\theta).
\end{eqnarray*}
Introduce these terms into $\E\var(U_i|p_1, \cdots, p_K)$, we have the final result 
\begin{eqnarray*}
	\mathrm{E}(\mathrm{Var}[U_i|p_1, \cdots, p_K] ) & = & \frac{\theta^2}{12} N_{i+1} (\sum_{j=1}^i N_j)(\sum_{j=1}^{i+1} N_j)(\sum_{j=1}^K N_j) [(\sum_{j=1}^K N_j)\theta + 3 - 2\theta]\\
	& \geq &  \frac{\theta^3}{12} N_{i+1} (\sum_{j=1}^i N_j)^2(\sum_{j=1}^K N_j)^2.
\end{eqnarray*}
The final inequality comes from $3 - 2\theta > 0$ and $\sum_{j=1}^{i+1} N_j > \sum_{j=1}^i N_j$. 

Therefore, the lemma is proved.

\section{Proof of Lemma \ref{lemma:diffcov}}
\setcounter{equation}{0}

We are interested in the case that $f_1 = f_2 = \cdots = f_K = f$, and all $\theta_i$'s are equal. For this case, we have that 
\begin{equation}\label{eqn:diffcovall}
	\cov(U_i, U_j) = \E[\cov(U_i, U_j |p_1, \cdots, p_K)] + \cov(\E[U_i|p_1, \cdots, p_K], \E[U_j|p_1, \cdots, p_K]) = I+II. 
\end{equation}
We want to find the upper bound for $I$ and $II$ separately. 

Consider $I$ first. Given $p_1, \cdots, p_K$, the number of non-zero elements are given, so the random part is the rank of non-zero elements in every sample. Let $r_i^0$ denote the sum-rank of the non-zero elements in sample $i$. Therefore, we calculate the conditional covariance as 
\begin{eqnarray*}
	\cov(U_i, U_j|p_1, \cdots, p_K) & = & \cov(N_{i+1} \sum_{l_1 = 1}^i r^0_{l_1} - r_{i+1}^0 \sum_{l_1 = 1}^i N_{l_1}, N_{j+1} \sum_{l_2 = 1}^j r^0_{l_2} - r_{j+1}^0 \sum_{l_2 = 1}^j N_{l_2})\\
	& = & \frac{1 + \sum_{k=1}^K n_k}{12}\biggl[ N_{i+1} N_{j+1}(\sum_{l_1 = 1}^i n_{l_1}) (\sum_{l_2 = j+1}^K n_{l_2}) + N_{i+1} n_{j+1}(\sum_{l_1 = 1}^i n_{l_1}) (\sum_{l_2 = 1}^j n_{l_2})\\
	&&- n_{i+1}N_{j+1} (\sum_{l_1 = 1}^i N_{l_1}) (\sum_{l_2 = j+1}^K n_{l_2}) 
	- n_{i+1} n_{j+1} (\sum_{l_1 = 1}^i N_{l_1}) (\sum_{l_2 = 1}^j N_{l_2}) \biggr].
\end{eqnarray*}
Note that $n_i \sim Binomial(N_i, \theta)$ independently, so that we have 
\begin{equation}\label{eqn:diffcovpart1}
	I = \E[\cov(U_i, U_j|p_1, \cdots, p_K)] = 0. 
\end{equation}

Next consider $II$. According to (\ref{eqn:eucond}), we have the result that 
\[
\mathrm{E}[U_i] =\frac{1}{2} N_{i+1} \sum_{k = 1}^{K} N_k \sum_{j = 1}^{i } N_j \mathrm{E}[p(p_{i+1} - p_j)] + 
O\biggl(\sum_{k = 1}^{K} N_k \sum_{j = 1}^{i + 1} N_j\biggr), \qquad 1 \leq i \leq K - 1,
\]
where the remainder comes from the difference between $\lfloor pN_i \rfloor$ and $p N_i$, which is a smaller order term. 
Therefore, the covariance between two terms is 
\begin{eqnarray*}
	II & = & \cov(\frac{1}{2} N_{i+1} \sum_{k = 1}^{K} N_k \sum_{l_1 = 1}^{i } N_{l_1} p(p_{i+1} - p_{l_1}) , \frac{1}{2} N_{j+1} \sum_{k = 1}^{K} N_k \sum_{l_2 = 1}^{j } N_{l_2} p(p_{j+1} - p_{l_2}) )\\
	& = & \frac{1}{4} N_{i+1} N_{j+1} \bigl( \sum_{k = 1}^{K} N_k\bigr)^2  \cov( \sum_{l_1 = 1}^{i } N_{l_1}p(p_{i+1} - p_{l_1}),  \sum_{l_2 = 1}^{j } N_{l_2} p(p_{j+1} - p_{l_2}) ).
\end{eqnarray*}
It's hard to find an exact value, so we work on an upper bound only. We rewrite $p = \theta + (p - \theta)$, and so the covariance part is the summation of four parts
\begin{eqnarray*}
	&&\cov( \sum_{l_1 = 1}^{i } N_{l_1}p(p_{i+1} - p_{l_1}),  \sum_{l_2 = 1}^{j } N_{l_2} p(p_{j+1} - p_{l_2}) )\\
	& = & \cov( \sum_{l_1 = 1}^{i } N_{l_1}\theta(p_{i+1} - p_{l_1}),  \sum_{l_2 = 1}^{j } N_{l_2} \theta(p_{j+1} - p_{l_2}) )+ \cov( \sum_{l_1 = 1}^{i } N_{l_1}(p - \theta)(p_{i+1} - p_{l_1}),  \sum_{l_2 = 1}^{j } N_{l_2} \theta(p_{j+1} - p_{l_2}) )\\
	& &+ \cov( \sum_{l_1 = 1}^{i } N_{l_1}\theta(p_{i+1} - p_{l_1}),  \sum_{l_2 = 1}^{j } N_{l_2} (p - \theta)(p_{j+1} - p_{l_2}) ) + \cov( \sum_{l_1 = 1}^{i } N_{l_1}(p - \theta)(p_{i+1} - p_{l_1}),  \sum_{l_2 = 1}^{j } N_{l_2} (p - \theta)(p_{j+1} - p_{l_2}) )\\
	& = & IIa + IIb + IIc + IId.
\end{eqnarray*}
Check the terms one by one. 
For $IIa$, note that $\cov(p_i, p_j) = 0$ when $i \neq j$ because of independence, so we have 
\begin{eqnarray}
	IIa & = & \theta^2 \cov( \sum_{l_1 = 1}^{i } N_{l_1}(p_{i+1} - p_{l_1}),  \sum_{l_2 = 1}^{j } N_{l_2} (p_{i+1} - p_{l_2}) ) \nonumber\\
	& = & \theta^2 \biggl[ \sum_{l=1}^i [N_l^2 \var(p_l)] - (\sum_{l=1}^i N_l) N_{i+1} \var(p_{i+1})\biggr]
	\nonumber\\
	& = & \theta^2 \biggl[ \sum_{l=1}^i N_l \theta(1 - \theta) - (\sum_{l=1}^i N_l) \theta(1 - \theta)\biggr] = 0.\label{eqn:diffcovpart2a}
\end{eqnarray}

The parts $IIb$ and $IIc$ are symmetric. We analyze one and the result for the other one can also be achieved. 
\begin{eqnarray}
	IIb & = & \theta \cov( \sum_{l_1 = 1}^{i } N_{l_1}(p - \theta)(p_{i+1} - p_{l_1}),  \sum_{l_2 = 1}^{j } N_{l_2} (p_{i+1} - p_{l_2}) ) \nonumber\\
	& \leq & \theta \sqrt{\var\bigl[ \sum_{l_1 = 1}^{i } N_{l_1}(p - \theta)(p_{i+1} - p_{l_1})\bigr] \var\bigl[\sum_{l_2 = 1}^{j } N_{l_2} (p_{i+1} - p_{l_2})\bigr]}.
	\label{eqn:diffcovpart2b1}
\end{eqnarray}
Then we check the two variances. Consider the first one. Note that $p = \max_{1 \leq i \leq K} p_i$, so $(p - \theta)^2 \leq \sum_{i=1}^K (p_i - \theta)^2$. So we have 
\begin{eqnarray}
	\var\bigl[ \sum_{l_1 = 1}^{i } N_{l_1}(p - \theta)(p_{i+1} - p_{l_1})\bigr] & \leq & 
	\E[(p - \theta)^2] \E[(\sum_{l_1 = 1}^{i } N_{l_1}(p_{i+1} - p_{l_1}))^2]\nonumber\\
	& \leq & \biggl[\sum_{k=1}^K \E[(p_k - \theta)^2] \biggr] \E[(\sum_{l_1 = 1}^{i } N_{l_1}(p_{i+1} - p_{l_1}))^2]\nonumber\\
	(\mbox{Note } \E[\sum_{l_1 = 1}^{i } N_{l_1}(p_{i+1} - p_{l_1})] = 0)
	& = &\biggl[\sum_{k=1}^K \var(p_k - \theta) \biggr] \var[\sum_{l_1 = 1}^{i } N_{l_1}(p_{i+1} - p_{l_1})]\nonumber\\
	& = & \theta^2(1 - \theta)^2 (\sum_{k=1}^K 1/N_k)  (\sum_{l=1}^i N_l) [1 + \sum_{l=1}^i N_l/N_{i+1}]\label{eqn:diffcov2b1}
\end{eqnarray}
For the second one, the calculation is straigtforward. 
\begin{equation}\label{eqn:diffcov2b2}
	\var\bigl[\sum_{l_2 = 1}^{j } N_{l_2} (p_{i+1} - p_{l_2})\bigr] = \theta(1 - \theta) (\sum_{l=1}^j N_l) [1 + \sum_{l=1}^j N_l/N_{j+1}].
\end{equation}
Combine (\ref{eqn:diffcov2b1}) -- (\ref{eqn:diffcov2b2}) with (\ref{eqn:diffcovpart2b1}), we have that 
\begin{eqnarray}
	IIb & \leq & \theta \sqrt{\theta^2(1 - \theta)^2 (\sum_{k=1}^K 1/N_k)  (\sum_{l=1}^i N_l) [1 + \sum_{l=1}^i N_l/N_{i+1}] \theta(1 - \theta) (\sum_{l=1}^j N_l) [1 + \sum_{l=1}^j N_l/N_{j+1}]}\nonumber\\
	& \leq & \theta^2 [1 + \sum_{l=1}^K N_l/N_{(1)}] \sqrt{ (\sum_{k=1}^K 1/N_k)(\sum_{l=1}^i N_l)  (\sum_{l=1}^j N_l) },
	\label{eqn:diffcovpart2b}
\end{eqnarray}
where $N_{(1)}$ is the minimum of $N_1, \cdots, N_K$. 

Similarly, for $IIc$, we have 
\begin{equation}\label{eqn:diffcovpart2c}
	IIc \leq \theta^2 [1 + \sum_{l=1}^K N_l/N_{(1)}] \sqrt{ (\sum_{k=1}^K 1/N_k)(\sum_{l=1}^i N_l)  (\sum_{l=1}^j N_l) }.
\end{equation}

Finally, we check $IId$. With the results above, we have the upper bound for $IId$ as 
\begin{eqnarray}
	IId & \leq & \sqrt{\var\bigl[ \sum_{l_1 = 1}^{i } N_{l_1}(p - \theta)(p_{i+1} - p_{l_1})\bigr] \var\bigl[\sum_{l_2 = 1}^{j } (p - \theta)N_{l_2} (p_{i+1} - p_{l_2})\bigr]}\nonumber \\
	& \leq & \sqrt{\theta^2(1 - \theta)^2 (\sum_{k=1}^K 1/N_k)  (\sum_{l=1}^i N_l) [1 + \sum_{l=1}^i N_l/N_{i+1}] \theta^2(1 - \theta)^2 (\sum_{k=1}^K 1/N_k)  (\sum_{l=1}^i N_l) [1 + \sum_{l=1}^i N_l/N_{i+1}]}\nonumber\\
	& \leq & \theta^2  [1 + \sum_{l=1}^K N_l/N_{(1)}] (\sum_{k=1}^K 1/N_k)\sqrt{ (\sum_{l=1}^i N_l)  (\sum_{l=1}^j N_l) }.\label{eqn:diffcovpart2d}
\end{eqnarray}

Finally, combine (\ref{eqn:diffcovpart2a}), (\ref{eqn:diffcovpart2b}), (\ref{eqn:diffcovpart2c}) and (\ref{eqn:diffcovpart2d}), we have that 
\begin{eqnarray}
	&& |\cov( \sum_{l_1 = 1}^{i } N_{l_1}p(p_{i+1} - p_{l_1}),  \sum_{l_2 = 1}^{j } N_{l_2} p(p_{j+1} - p_{l_2}) )| =  |IIa + IIb + IIc + IId| \nonumber\\
	&\leq & 0 + \theta^2 [1 + \sum_{l=1}^K N_l/N_{(1)}] \sqrt{ (\sum_{k=1}^K 1/N_k)(\sum_{l=1}^i N_l)  (\sum_{l=1}^j N_l) } + 
	\theta^2 [1 + \sum_{l=1}^K N_l/N_{(1)}] \sqrt{ (\sum_{k=1}^K 1/N_k)(\sum_{l=1}^i N_l)  (\sum_{l=1}^j N_l) }\nonumber\\
	&&+ \theta^2  [1 + \sum_{l=1}^K N_l/N_{(1)}] (\sum_{k=1}^K 1/N_k)\sqrt{ (\sum_{l=1}^i N_l)  (\sum_{l=1}^j N_l) } \nonumber\\
	& \leq & 3\theta^2 [1 + \sum_{l=1}^K N_l/N_{(1)}] \sqrt{ (\sum_{k=1}^K 1/N_k)(\sum_{l=1}^i N_l)  (\sum_{l=1}^j N_l) },
	\label{eqn:diffcovpart2}
\end{eqnarray}
when $N_1, \cdots, N_K \rightarrow \infty$. 

Finally, combine (\ref{eqn:diffcovpart1}) and (\ref{eqn:diffcovpart2}) with (\ref{eqn:diffcovall}), there is 
\begin{eqnarray*}
	|\cov(U_i, U_j)| & = & |I + II| = |II| \leq \frac{1}{4} N_{i+1} N_{j+1} \bigl( \sum_{k = 1}^{K} N_k\bigr)^2 \cdot 3\theta^2 [1 + \sum_{l=1}^K N_l/N_{(1)}] \sqrt{ (\sum_{k=1}^K 1/N_k)(\sum_{l=1}^i N_l)  (\sum_{l=1}^j N_l) }\\
	& \leq &  \theta^2 N_{i+1} N_{j+1} \bigl( \sum_{k = 1}^{K} N_k\bigr)^2 \ [1 + \sum_{l=1}^K N_l/N_{(1)}] \sqrt{ (\sum_{k=1}^K 1/N_k)(\sum_{l=1}^i N_l)  (\sum_{l=1}^j N_l) }.
\end{eqnarray*}
Therefore, the result is proved.

 \end{document}